\documentclass[12pt,preprint]{aastex}

\begin{document}

\title{Radio Observations of the Hubble Deep Field South Region III: The 2.5,
5.2 and 8.7 GHz Catalogues and Radio Source Properties.}
\author{Minh T. Huynh} 
\affil{Spitzer Science Center, MC220-6, California Institute of Technology, Pasadena CA 91125, USA}
\email{mhuynh@ipac.caltech.edu}
\author{Carole A. Jackson} 
\affil{Australia Telescope National Facility, CSIRO Radiophysics Laboratory, PO 
Box 76, Epping, 
NSW 2121, Australia}
\author{Ray P. Norris}
\affil{Australia Telescope National Facility, CSIRO Radiophysics Laboratory, PO 
Box 76, Epping, 
NSW 2121, Australia}

\begin{abstract}
Deep radio observations of a wide region centred  on the Hubble Deep Field South have
been performed, providing one of the most sensitive set of radio observations
acquired on the Australia Telescope Compact Array to date. A central rms of $\sim$$10\mu$Jy is 
reached at four frequencies (1.4, 2.5, 5.2 and 8.7
GHz). In this paper the full source catalogues from the
2.5, 5.2 and 8.7 GHz observations are presented to complement Paper II, 
along with a detailed analysis of image quality and noise. We produce a 
consolidated catalogue by matching sources
across all four frequencies of our survey. Radio spectral indices are used to investigate 
the nature of the radio sources and identify a number of sources with flat or 
inverted radio spectra, which indicates AGN activity. We also find several other interesting sources, including a 
broadline emitting radio galaxy, a giant radio galaxy and three Gigahertz Peaked Spectrum sources.

\end{abstract}


\section{Introduction}

Deep radio surveys over the last decade have pointed to the presence of a new
population of radio sources distinct from classical radio galaxies and
QSOs. These sources start to dominate the count for $S_{\rm 1 GHz} \lesssim 1$ mJy and they
are responsible for the flattening in the source count for sub-mJy
levels. Many different schemes have been put forward to explain this `excess' of faint
radio sources, including a non-evolving population of local ($z < 0.1$) low
luminosity radio galaxies \citep{wall86}, strongly evolving normal spirals
\citep{condon84, condon89} and starburst galaxies \citep{windhorst85,rowan93}. 
The sub-mJy radio source population is now thought to be a mix of low luminosity AGN, normal spirals
and ellipticals, as well as starbursts, but surprisingly little is known about the
exact nature of the faint radio population and the sources which dominate. 
This is because sub-mJy radio sources have faint optical counterparts and hence 
obtaining complete optical identifications has been historically difficult.

To help alleviate the significant time overheads of optical followup, we have targeted
the Hubble Deep Field South (HDFS). This is a very well-studied field with publicly
available data ranging in wavelengths from radio to UV-optical. The Hubble
Space Telescope has observed this field to about 30th magnitude in the optical 
\citep{casertano00}, and similar depths in UV \citep{gardner00} and NIR \citep{yahata00}.
In addition to the main deep fields, the HST imaged nine Flanking Fields to a
depth of I(F814W) $\sim$ 26. From the ground, wide field imaging is available
for an area $44 \times 44$ arcmin in extent around the HDFS, which reaches
about 25th magnitude in UBVRI \citep{teplitz01}. Spectroscopy has been obtained on the VLT for
194 targets in the main HDFS and Flanking Fields, resulting in reliable
redshifts for 97 galaxies \citep{sawicki03}. Photometric redshifts are also available for thousands of galaxies in the HDFS and surrounding regions \citep{teplitz01, rudnick2001, labbe2003}, which have typical errors $\delta z / (1 + z) \lesssim 0.1$. 

Radio observations of the HDFS were made between 1998 and 2001 with the
Australia Telescope Compact Array (ATCA) using all four available frequency
bands. Between 100 to 300 hours of observing at each band yielded images at
1.4, 2.5, 5.2 and 8.7 GHz with maximum sensitivities of $\sim 10\mu$Jy rms. 
A detailed description of the observations, data reduction, and initial results 
was given by \cite{norris05} (hereafter Paper I). The full 1.4 GHz
catalogue, a detailed 1.4 GHz image analysis and radio source count was 
presented in \cite{huynh05} (hereafter Paper II). 

The advantage of the ATHDFS dataset over other deep radio surveys is that it is at 
four frequencies, making it possible to do an analysis
of the radio spectra of the individual sources and identify those with spectra characteristic of AGN. 
This paper details the 2.5, 5.2 and 8.7 GHz catalogue and an analysis of the radio spectral properties of sources in the ATHDFS. 

This paper is outlined as follows. We summarise the the 2.5, 5.2 and 8.7 GHz
observations and data reduction in Section \ref{sec:obs}. 
A detailed analysis of the images at these three frequencies is discussed in Section
\ref{imageanalysis}, and the catalogues listed in Section \ref{catalogues}. We
present a consolidated catalogue in Section \ref{combcat} where radio sources have
been matched across all four ATHDFS bands. Section \ref{alphasect} presents an analysis of
the radio spectral index of ATHDFS sources and other interesting radio sources are discussed in Section \ref{inter_src}.

We assume a Hubble constant of 71 km s$^{-1}$ Mpc$^{-1}$, $\Omega_{\rm M}  = 0.27$ and $\Omega_{\rm \Lambda}  = 0.73$ throughout this paper. 

\section{Observations and Data Reduction}
\label{sec:obs}

Here we provide a brief summary of the 2.5, 5.2 and 8.7 GHz observations and data reduction 
which are  discussed in detail in Paper I. The observations consist of single pointings 
centred on RA = 22h 33m 25.96s and Dec = $-$60$^\circ$ 38' 09.0'' (J2000) (2.5
GHz), and RA = 22h 32m 56.22s and Dec = $-$60$^\circ$ 33' 02.7'' (J2000) (5.2 and
8.7 GHz). The 5.2 and 8.7 GHz observations are centred on the HST WFPC field,
while the 2.5 GHz observations were pointed halfway between the WFPC
field and a bright confusing source to allow the bright source to be well 
cleaned from the 2.5 GHz image.

We used a wide variety of ATCA configurations to maximise {\em uv} coverage. 
The correlator was set to continuum mode ($2 \times 128$ MHz
bandwidth), with each 128 MHz bandwidth divided into $32 \times 4$ MHz channels.
The primary flux density calibrator is PKS B1934-638, while 
secondary gain and phase calibrations were taken throughout our observations 
using both PKS B2205-636 and PKS B2333-528. 

The 2.5 and 5.2 GHz images contained sidelobes from off-field sources. The
image sensitivity was improved by removing the clean components of these 
off-field sources before the final imaging. The 2.5 GHz image was also improved with one
iteration of phase and amplitude self-calibration. The 5.2 and 8.7 GHz data
were of sufficient quality that self-calibration was not needed. 

\section{Analysis of 2.5, 5.2 and 8.7 GHz Images}
\label{imageanalysis}

The parameters of the final ATHDFS images are summarised in Table \ref{athdfssummarytab}. The 1.4 GHz values are included for completeness.
For each frequency we list the pointing (J2000), number of effective hours observed, and synthesised beam 
(size and position angle). The rms noise at the image centre is also given, along with the value of the most negative pixel. 

\subsection{Clean Bias}

If {\em uv} coverage is poor, the cleaning process can redistribute flux from
real sources on to noise peaks. Our {\em uv} coverage is good, and in Paper II
we found that clean bias was small at 1.4 GHz. Here we perform a similar
test on the 2.5, 5.2 and 8.7 GHz images. 

To perform the clean bias check, point sources are injected random positions
in the data before cleaning, and the data cleaned to the same level as the
final images. Source peak fluxes before and after cleaning were then
compared. We injected 40 sources at $5 \sigma$, 15 at 
$6 \sigma$, 15 at $7\sigma$, 15 at $8\sigma$, 15 at $9\sigma$, 10 at $10\sigma$,
3 at $20 \sigma$, 2 at $30 \sigma$, 1 at $50 \sigma$ and 1 at $100
\sigma$. This simulation was repeated 50 times to get reliable number statistics. 

\begin{figure}
\centering
     \includegraphics[width = 7cm]{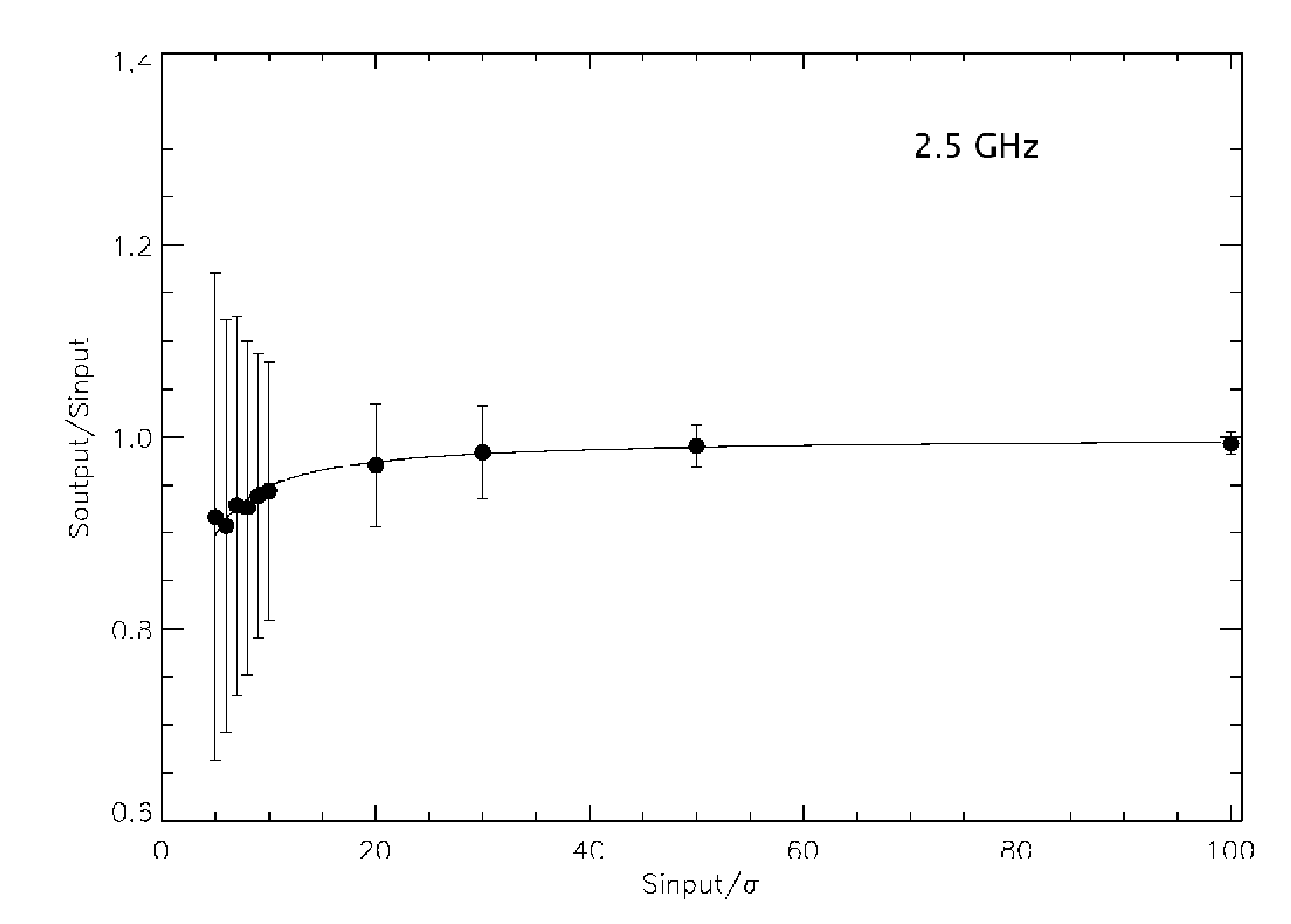}

     \includegraphics[width = 7cm]{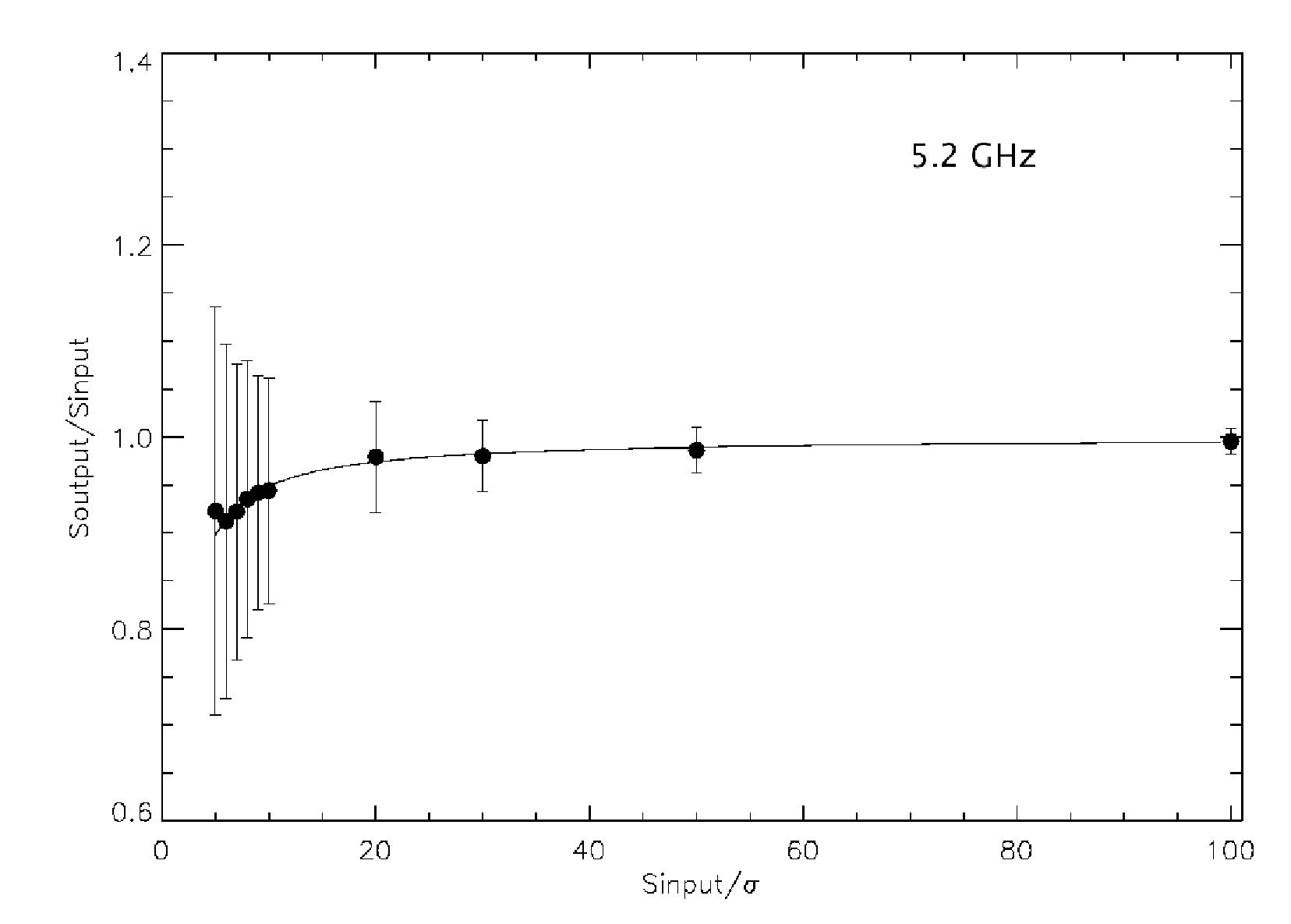}

     \includegraphics[width = 7cm]{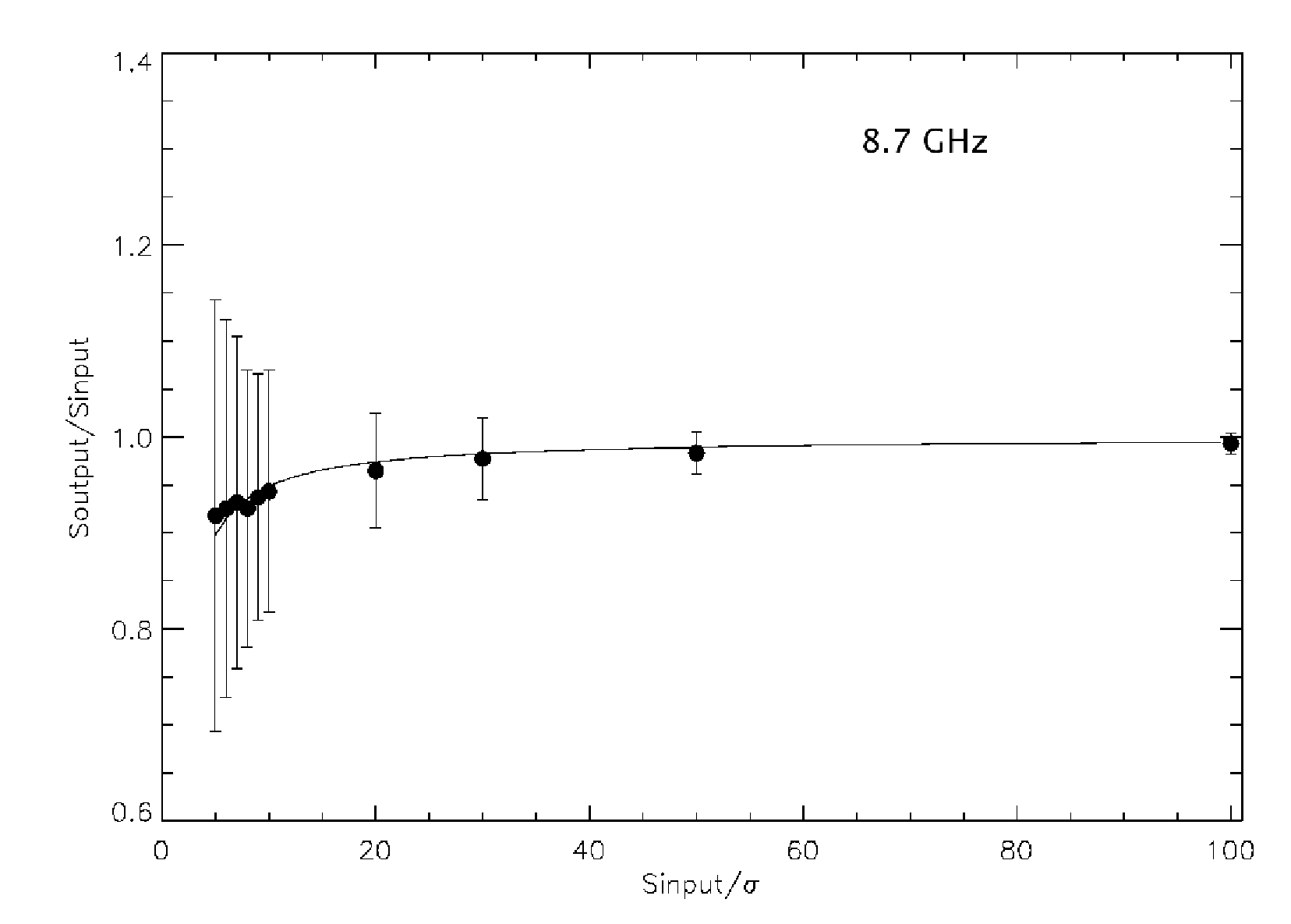}
\caption{Source fluxes measured after the cleaning process (S$_{\rm output}$) 
normalised to the true source fluxes (S$_{\rm input}$), as a function of the 
input source signal-to-noise (S$_{\rm input}/\sigma_{\rm local}$). 
Also shown are the best fit curves (see Section 3.1).}
\label{fig:clnbias}
\end{figure}

Figure \ref{fig:clnbias} shows the result of our clean bias check. The average 
source flux measured after the cleaning process (S$_{\rm output}$) divided by
the true source flux (S$_{\rm input}$) is shown for the various values of input 
source signal-to-noise (S$_{\rm input}/\sigma_{\rm local}$). 
It is evident that clean bias only affects the faintest sources.

An analytical form of the clean bias effect was obtained by a least squares 
fit to the function \[ S_{\rm output}/S_{\rm input} = a + b 
\frac{1}{S_{\rm input}/\sigma_{\rm local}} .\,\] 
The best fit values for $a$ and $b$ are shown in Table \ref{clnbiasfit}. 
The resulting best fit curves are plotted as solid lines in Figure
\ref{fig:clnbias}.

\subsection{Bandwidth Smearing}
\label{sec:bandsmear}

Bandwidth smearing is a well-known effect caused by the finite width of the
receiver channels. This aberration reduces the peak flux density of a source
while correspondingly increasing the size of a source such that the integrated
flux density is conserved. In Paper II we found that the 1.4 GHz image is
affected by bandwidth smearing, and at 20 arcmin from the phase centre, which
is the 1.4 GHz catalogue limit, sources are attenuated by $\sim$18\%. 

\begin{figure}

\centering

     \includegraphics[width = 7cm]{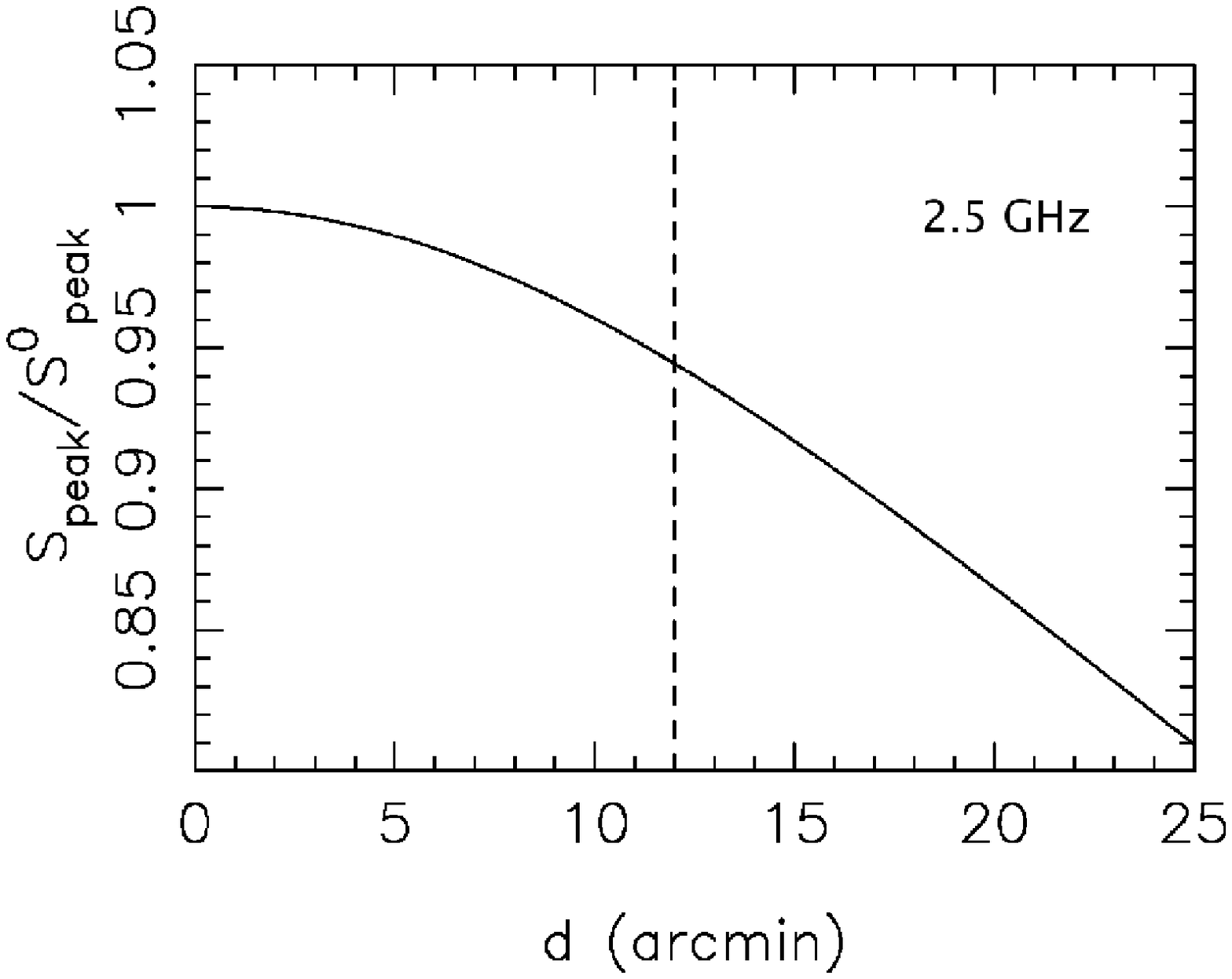}

     \includegraphics[width = 7cm]{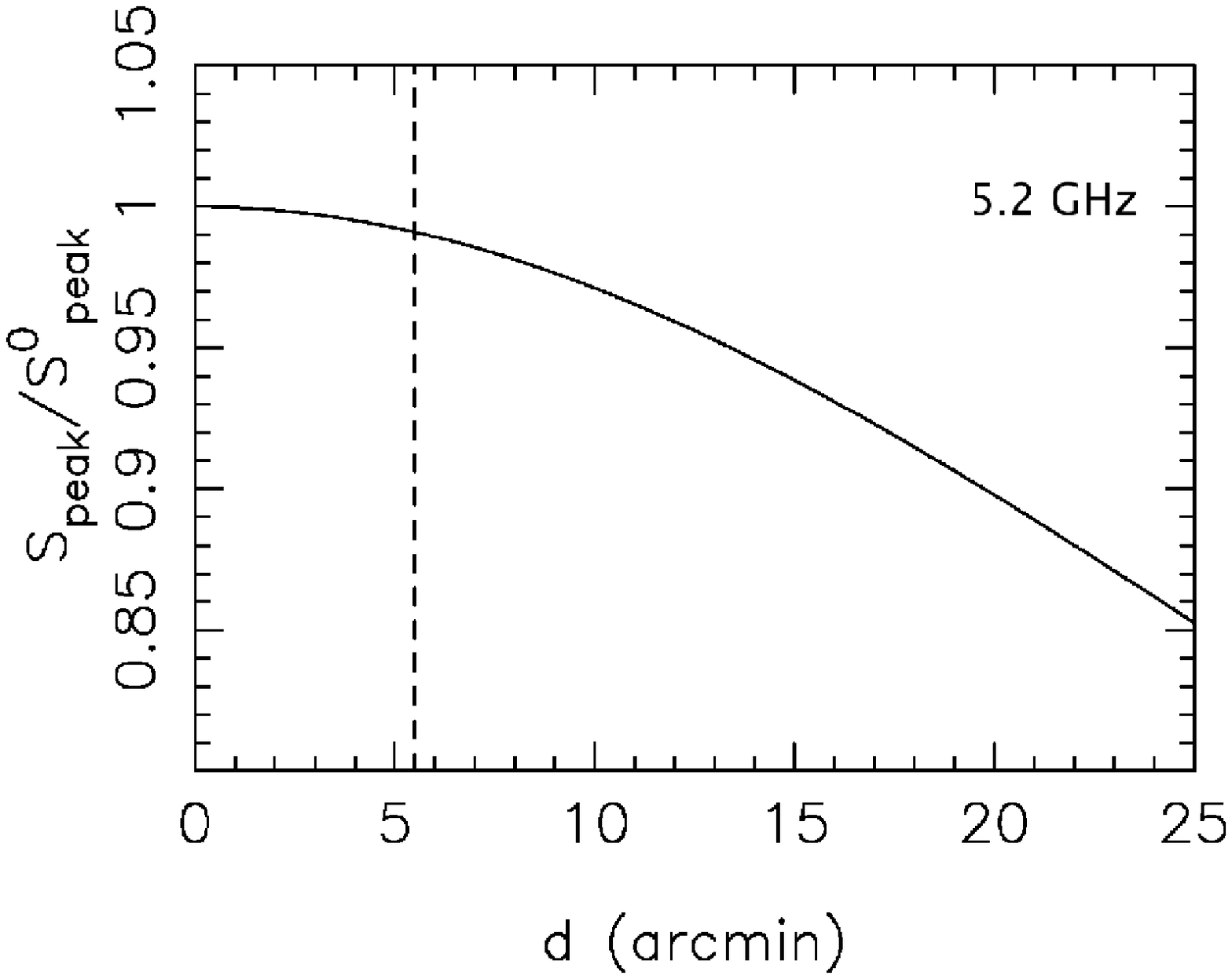}

     \includegraphics[width = 7cm]{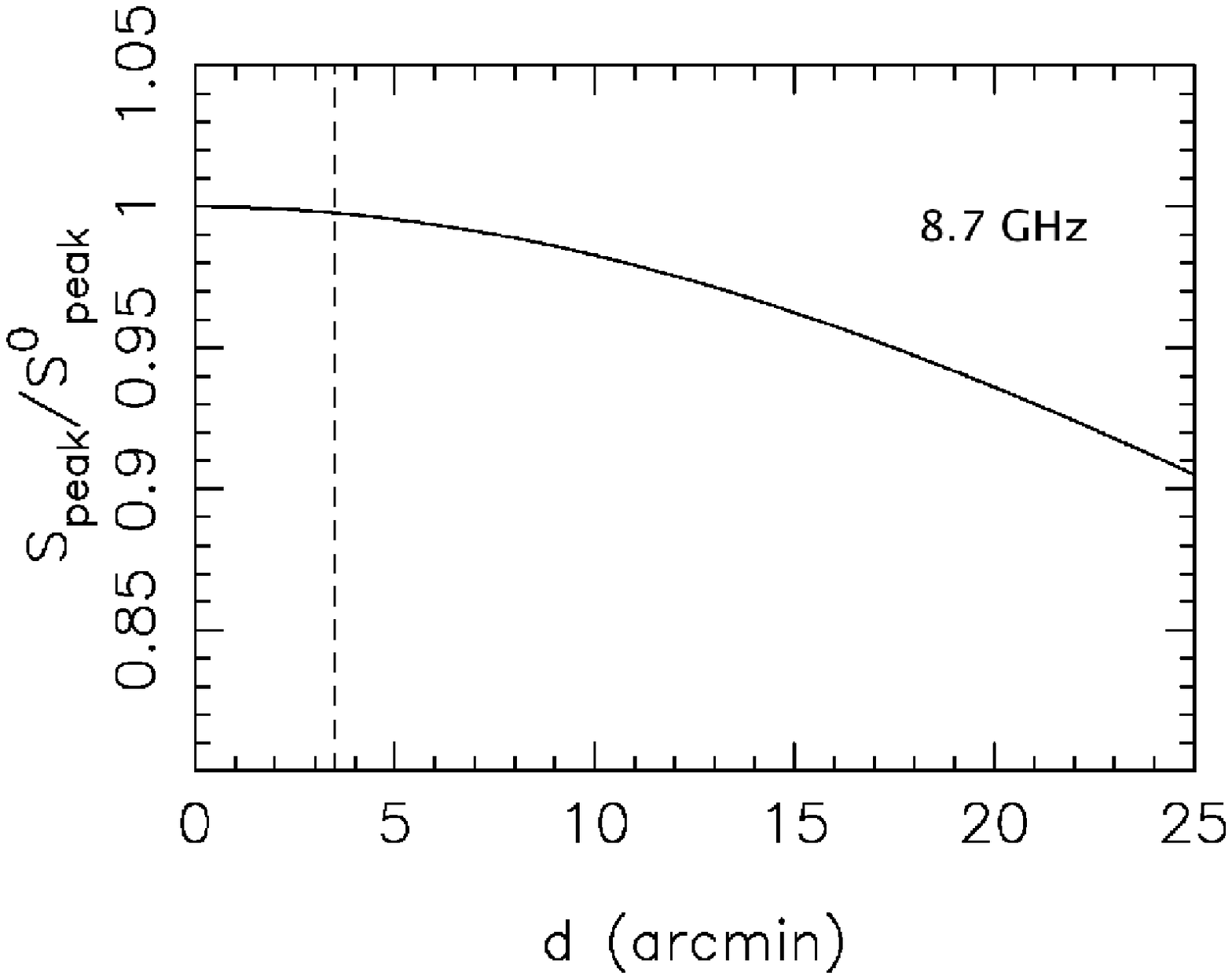}
\caption{Bandwidth smearing effect, as a function of distance from the phase centre, for 
three ATHDFS images. The dashed lines indicate the maximum distance to which sources 
are catalogued for each image (see Section 3.2 for details).}
\label{fig:bandsmear}
\end{figure}

We calculate the bandwidth smearing attenuation for the 2.5, 5.2 and 8.7 GHz
images using the appropriate beam parameters and Equation 1 from
Paper II. Since bandwidth smearing is inversely proportional to frequency we
expect the attentuation to be lower at these higher frequencies than at 1.4
GHz. The bandwidth smearing as a function of distance from the centre for the
three images are shown in Figure \ref{fig:bandsmear}. The dashed lines in
Figure \ref{fig:bandsmear} indicate the maximum distance at which sources are catalogued. 
We find bandwidth smearing is at most 5\% in the 2.5 GHz image, and 
negligible at the two higher frequencies.

\subsection{Noise Maps}

To investigate the noise characteristics of our images, and eventually 
obtain catalogues based on the local signal to noise ratio (SNR), 
we construct noise maps. Noise maps which contain the pixel by pixel 
root mean square (rms) noise distribution of each of the ATHDFS images 
were made using the SExtractor package \citep{bertin96}. SExtractor has been found to
be to be reliable for radio images (Paper II, \citealp{bondi03}).

\begin{figure}

\centering

     \includegraphics[width = 6.4cm]{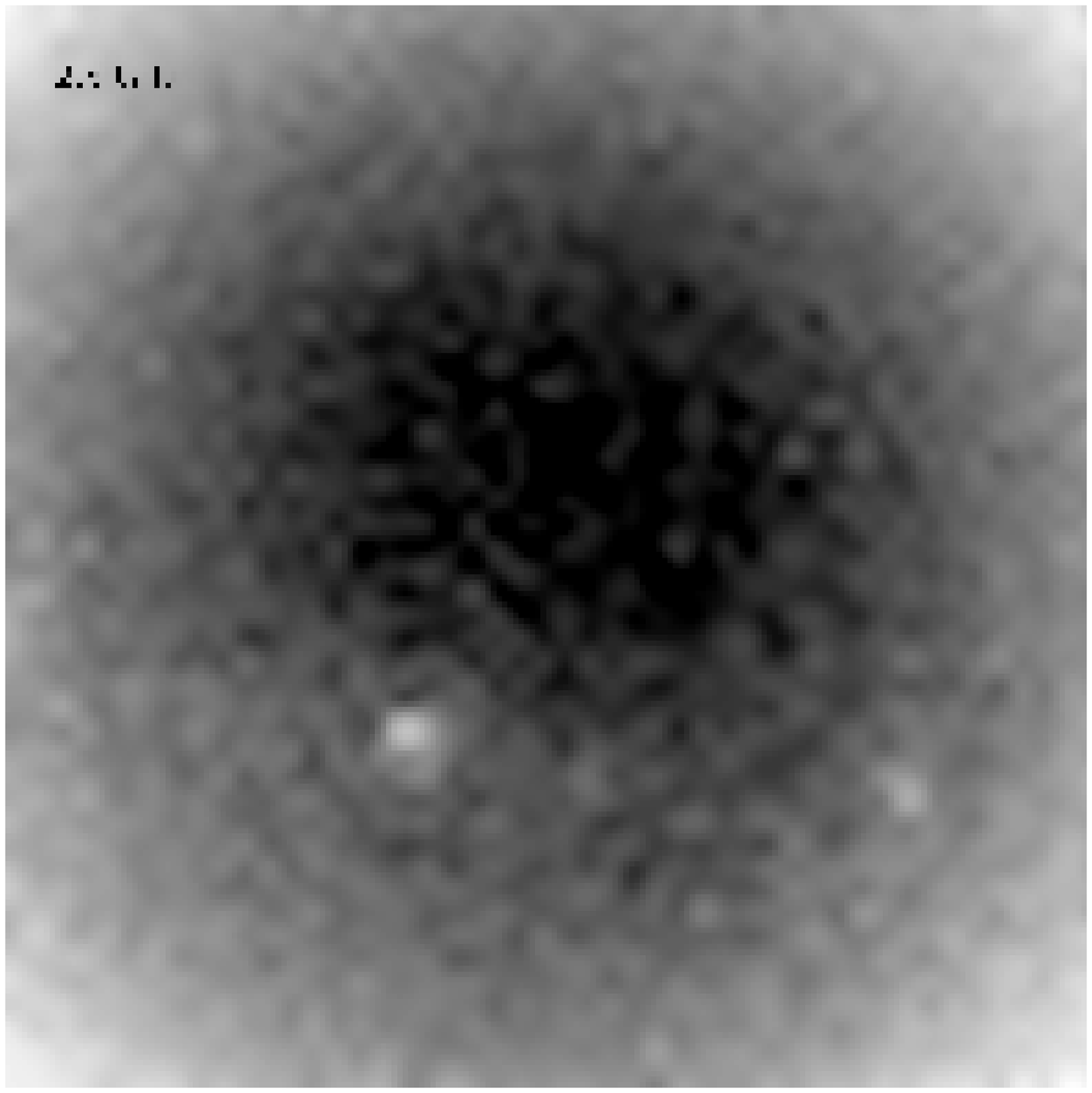}

     \includegraphics[width = 7cm]{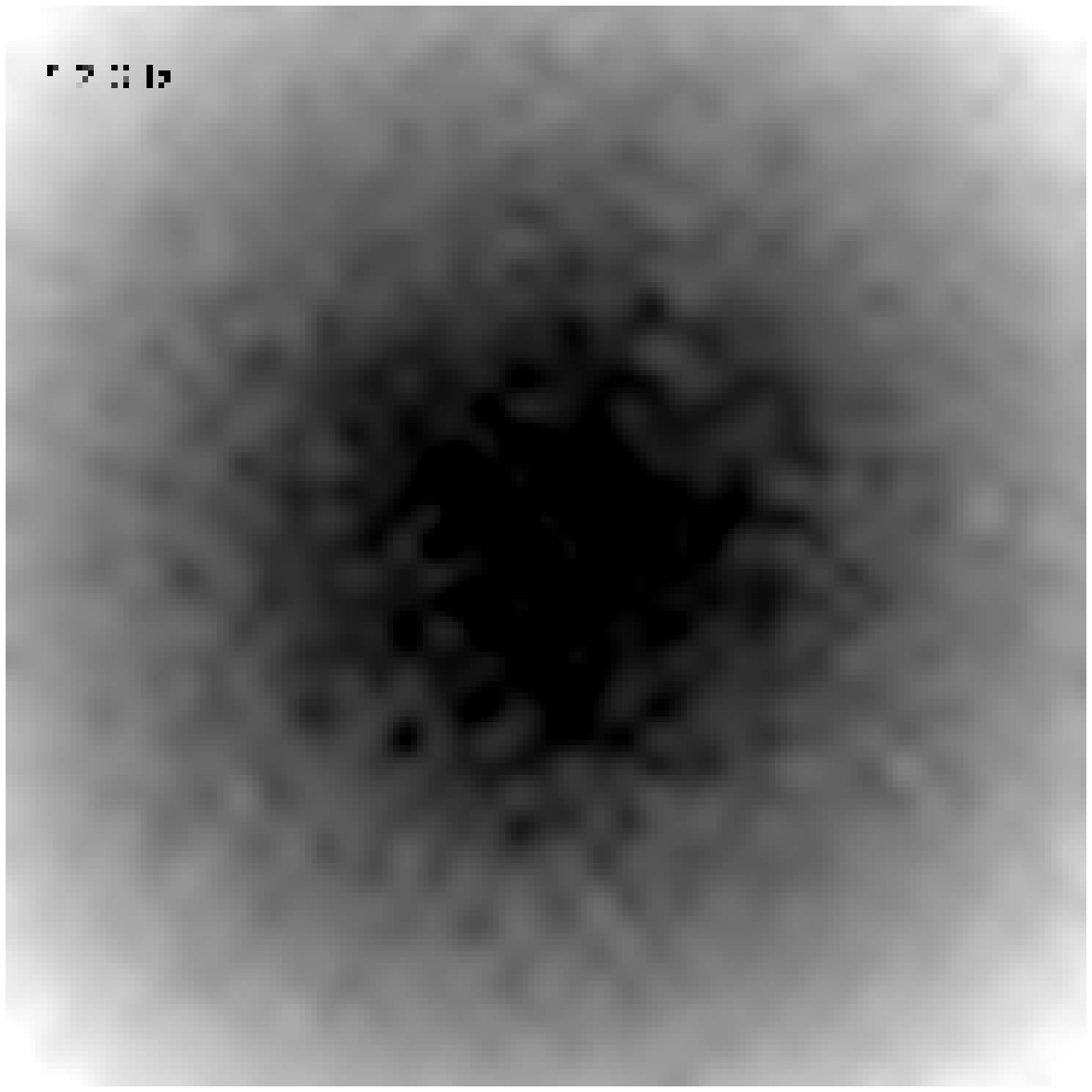}

     \includegraphics[width = 7cm]{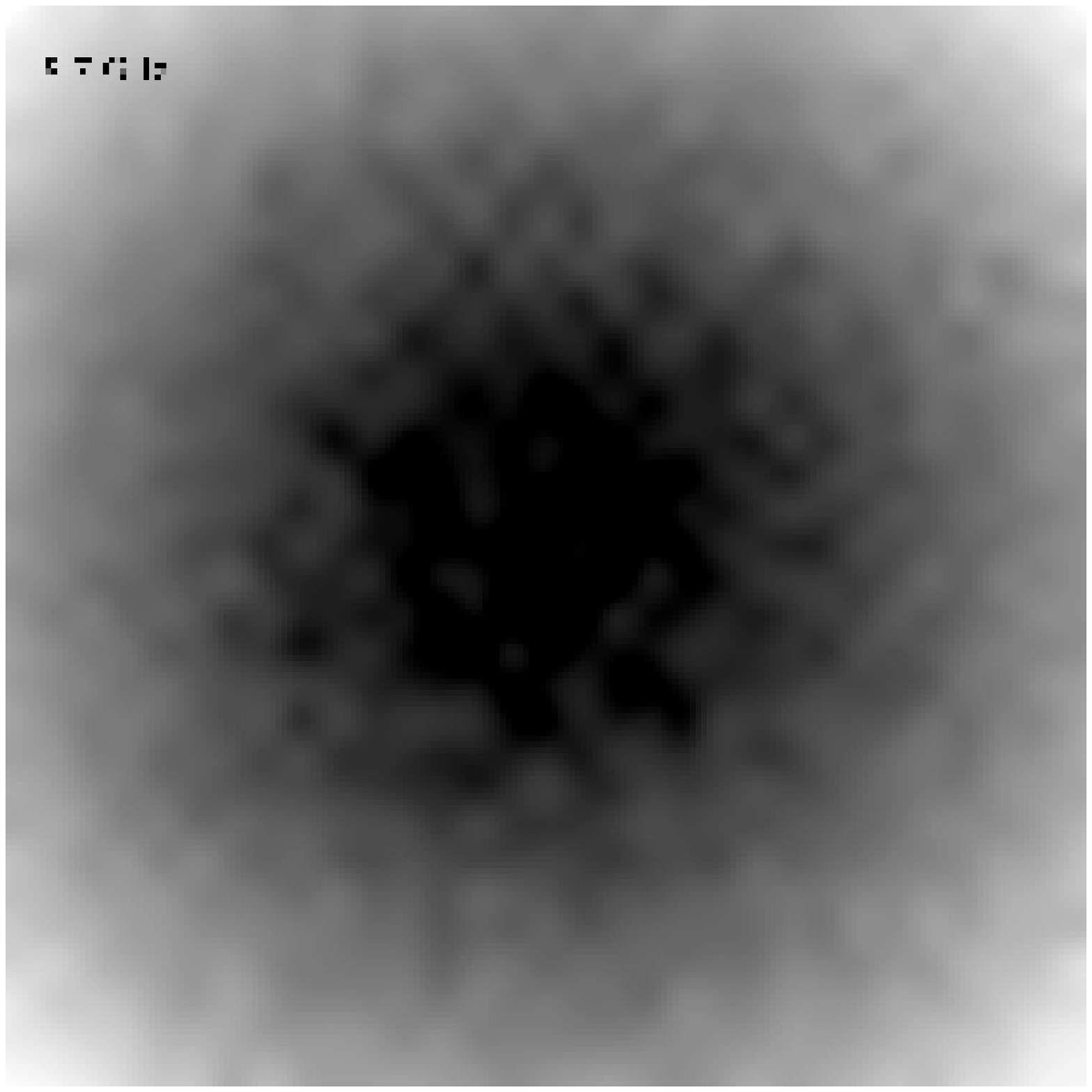}
\caption{Grey scales of the noise maps obtained by SExtractor. The images are 
28$\times$ 28 arcmin, 12$\times$ 12 arcmin, and 
8$\times$ 8 arcmin, for 2.5, 5.2 and 8.7 GHz, respectively.
The darker regions indicate lower noise.}
\label{noisemap}
\end{figure}

\begin{figure}
      \includegraphics[width = 8cm]{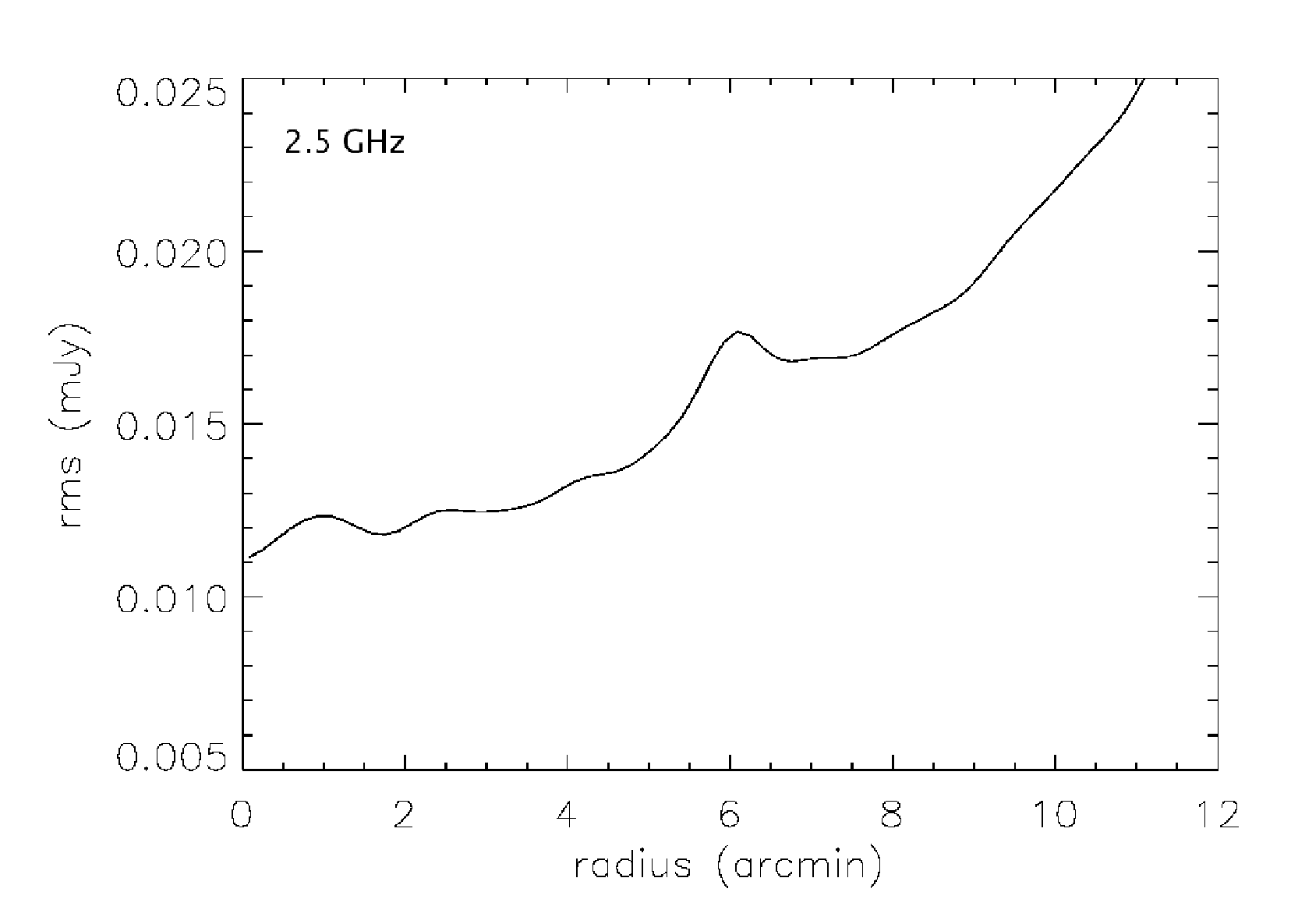}
      \includegraphics[width = 8cm]{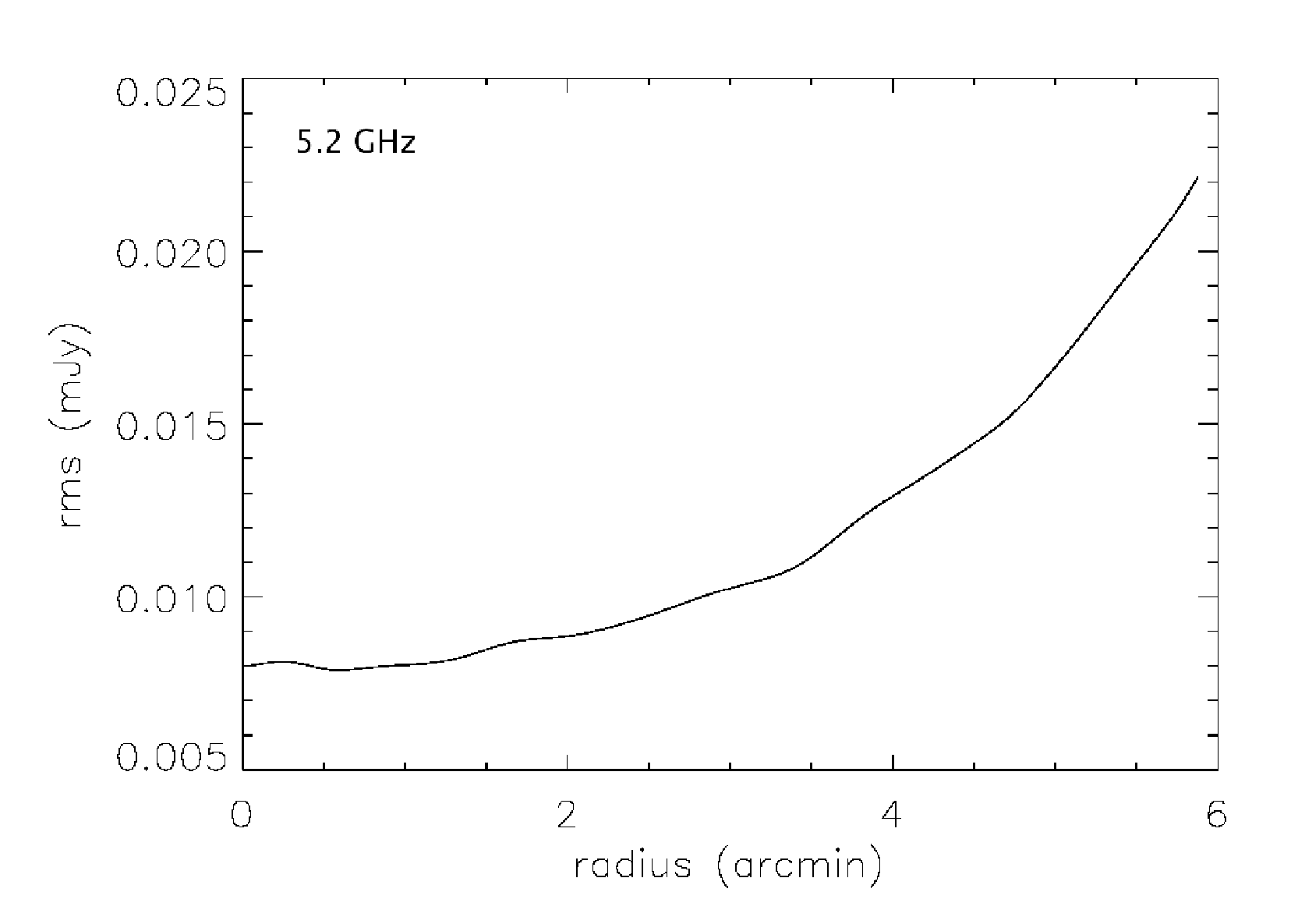}
      \includegraphics[width = 8cm]{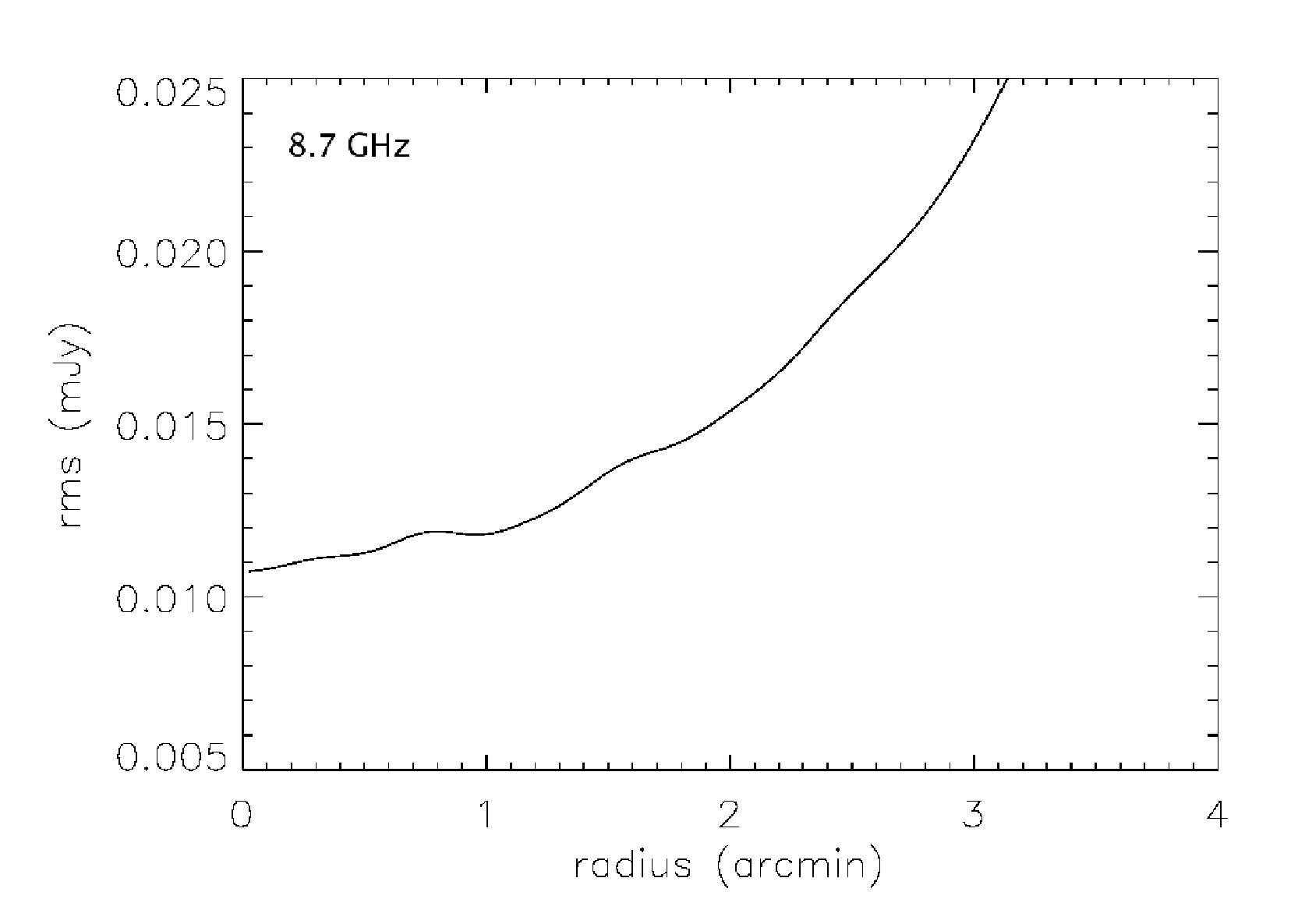}
\caption{Noise (radially averaged) as a function of radial distance for the SExtractor 
noise maps of the ATHDFS images.}
\label{radialnoiseprops}
\end{figure}

\begin{figure}

\centering

      \includegraphics[width = 7cm]{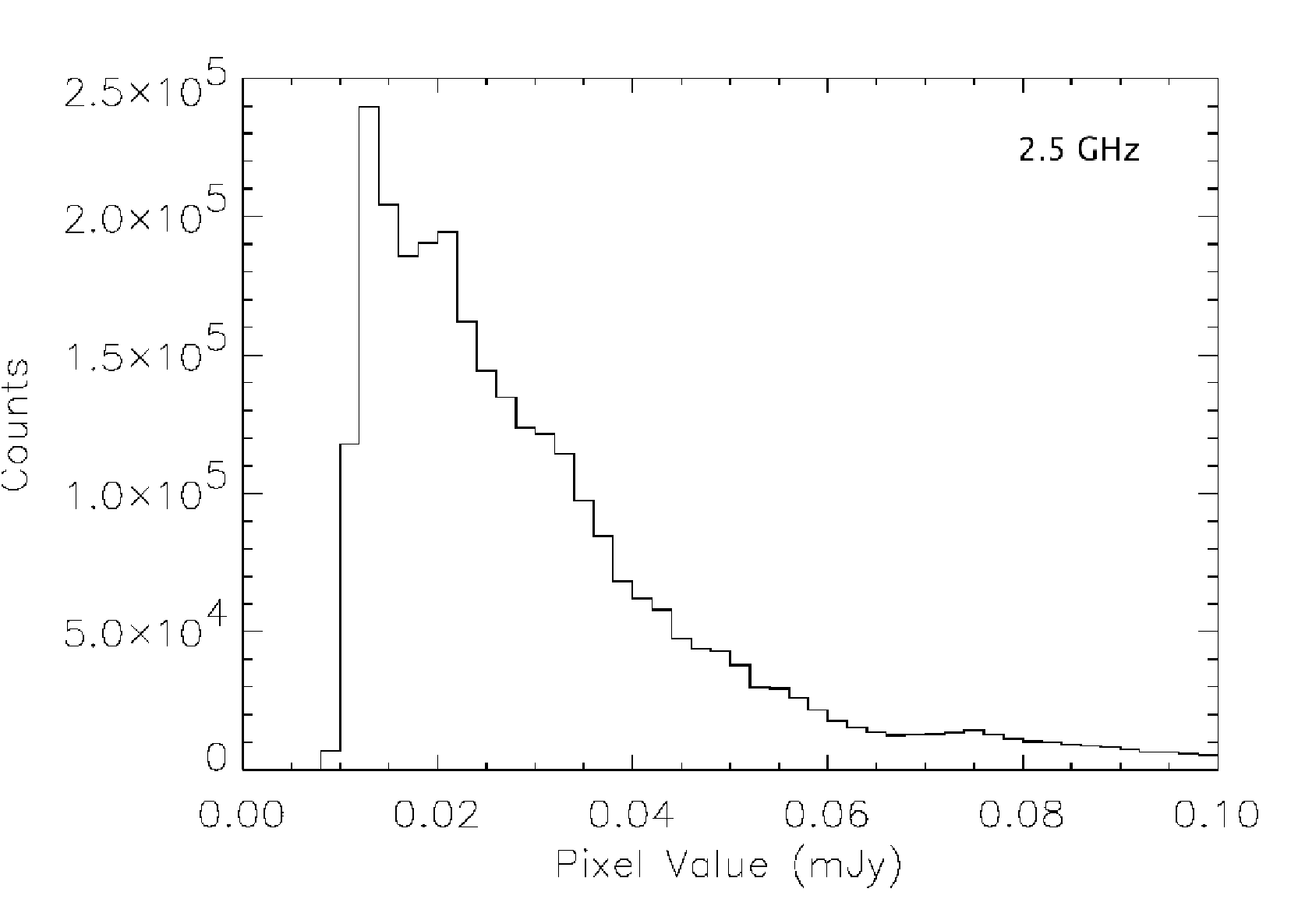}

      \includegraphics[width = 7cm]{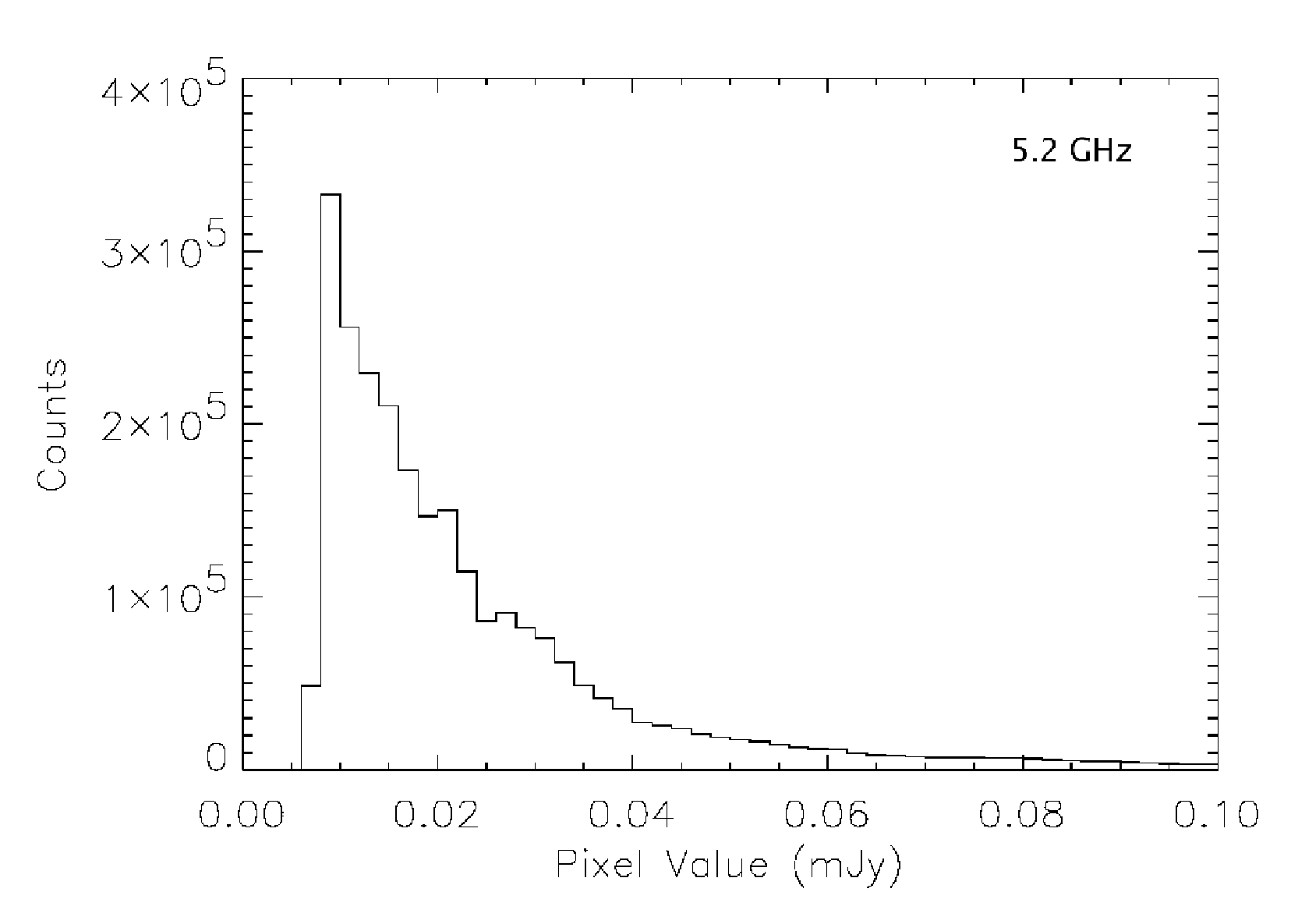}

      \includegraphics[width = 7cm]{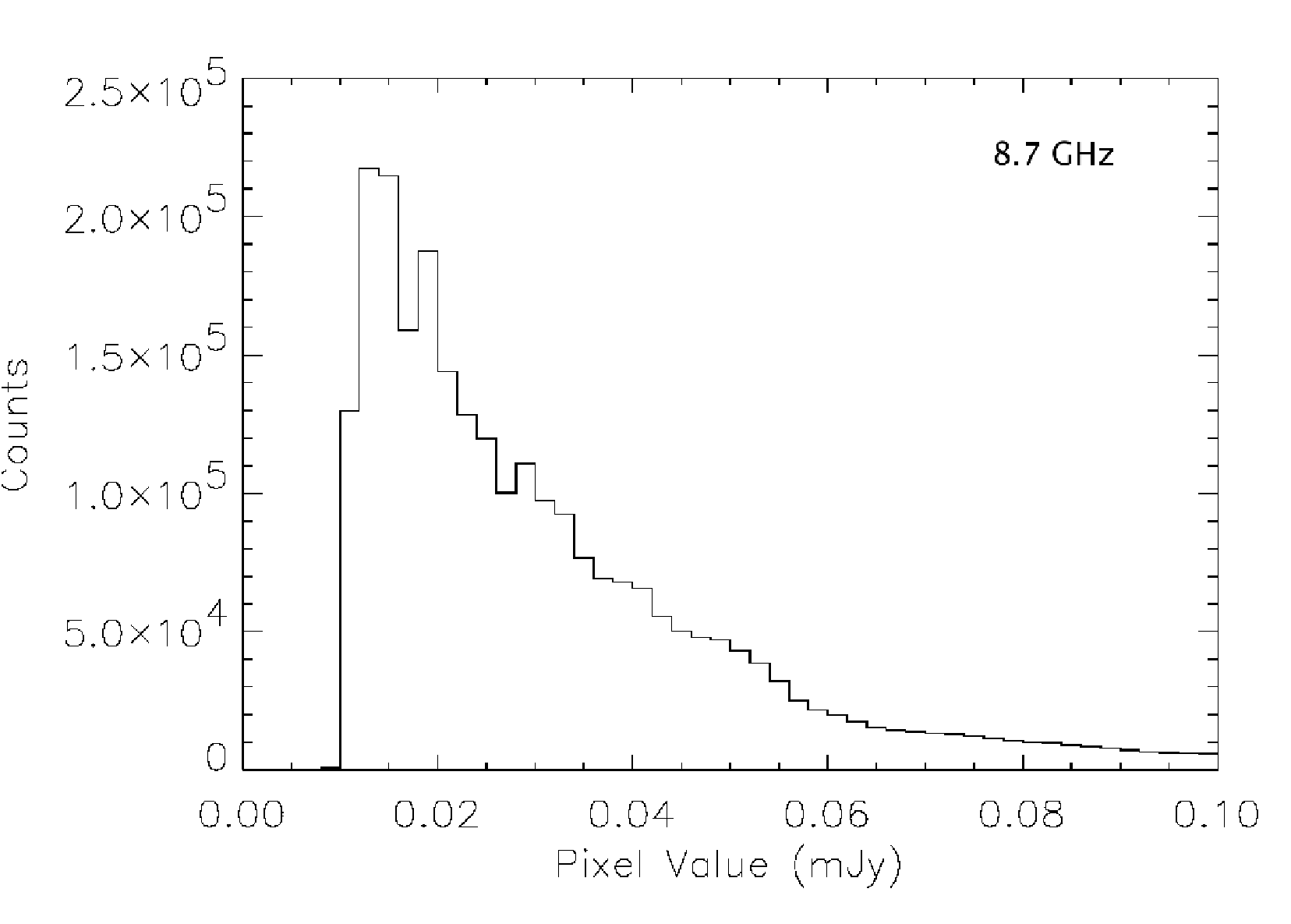}
\caption{Distribution of the pixel values of the SExtractor noise maps.}
\label{histnoiseprops}
\end{figure}

As for Paper II, we ran SExtractor on the ATHDFS images with a mesh size set to 8 $\times$ 8
beams. Grey scale images of the SExtractor noise maps are shown in Figure \ref{noisemap}. 
The noise is lowest in the centre and increases with radial distance, as
expected from single-pointing observations which are dominated by primary beam effects.
Figure \ref{radialnoiseprops} shows the average noise as a function of distance from the centre 
for the three ATHDFS noise maps. This increase in noise is roughly a parabolic shape due to 
the primary beam attenuation. Histograms of the pixel values of each noise map are shown 
in the Figure \ref{histnoiseprops}. The distributions peak at a value of about
10 $\mu$Jy, but have large wings at high noise values due to the radially increasing noise.

In Paper II we found regions of increased noise around bright sources. The 2.5
GHz noise map clearly shows increased noise around the brightest ($S_{\rm
2.5\,GHz}$ = 112 mJy) source, although the dynamic range problem is not as large here as for 1.4 GHz.
The area around this bright confusing source has noise which is up to a factor
of 2 greater than the nearby unaffected region. This can be seen as the noise
bump at approximately 6~arcmin from the 2.5 GHz image centre. The noise in the
5.2 and 8.7 GHz images is unaffected by bright sources. 

\section{Source Extraction at 2.5, 5.2 and 8.7 GHz} 
\label{srcext}

A sensible maximum radial distance for the source cataloging has been
determined, accounting for the following factors: i) the radial noise distribution, ii) the primary beam attenuation, and  iii) the bandwidth smearing effect.  
The dominant effect at 2.5, 5.2 and 8.7 GHz is the primary beam
attenuation. Similar to Paper II for 1.4 GHz, we catalogue to the 40\%
response level of the primary beam. Thus the maximum radial distances are 
12, 5.5 and 3.5 arcmins for the 2.5, 5.2 and 8.7 GHz images, respectively. As
mentioned in Section \ref{sec:bandsmear}, this results in bandwidth smearing
of only 5\% {\em at worst} for sources at the edge of the 2.5 GHz catalogued region.

Signal-to-noise maps are generated by dividing the original radio maps by the
SExtractor noise map. 
The {\it MIRIAD} task {\it IMSAD} was then used to derive a preliminary 
list of source ``islands'' with a peak flux density level of at least
$4\sigma$. This task searches for ``islands'' of pixels above a cutoff, set
here to $4\sigma$, and attempts to fit gaussian components to the ``islands''. 
This resulted in 258, 80 and 36 source ``islands'' 
extracted at 2.5, 5.2 and 8.7 GHz, respectively.

To derive source flux densities and sizes, we fit each source ``island'' found by {\it IMSAD} with an elliptical Gaussian. All sources were visually inspected for obvious failures and poor fits. A reference peak value was derived using the ({\it MIRIAD} interpolation task {\it maxfit}), and the Gaussian fit was considered good if the difference between the fitted peak and reference peak was less than 20\% of the reference value and the fitted position was inside the 0.9$S_{\rm peak}$ flux density contour. We compared the Gaussian integrated fluxes to fluxes directly measured from summing pixels greater than $3\sigma$ in the source area. In many cases the Gaussian fit provided good values for the position and peak flux densities but not for integrated flux densities. These sources are flagged with ``2'' in the catalogue. 

The reliability of the sources was estimated by running source extraction on the negative images. This is a valid estimate provided the cleaning process didn't add extra negative or positive peaks on to the images. As described in Paper I, the clean model residuals were monitored and the cleaning process was stopped when there was no reduction from extra components. So we expect that the cleaning process didn't add many positive or negative peaks. 
Table \ref{reliability} summarizes the results of running source extraction on the negative image. At 5$\sigma$ the 5.2 and 8.7 GHz catalogues have over 96\% reliability.   At 2.5 GHz we have enough statistics to examine the 5 -- 5.5 $\sigma$ sources, and find that these are only about 40\% reliable. With a SNR greater than 5.5 $\sigma$ the 2.5 GHz catalogue would have about 99\% reliability. We thus cut off the catalogues at 5.5$\sigma$, 5$\sigma$ and 5$\sigma$ at 2.5, 5.2 and 8.7 GHz. The final catalogues have 71, 24 and 6 sources at 2.5, 5.2 and 8.7 GHz, respectively.

Given a prior 1.4 GHz position, it maybe feasible to push the detection limit lower than 5$\sigma$. We searched for low SNR sources by matching 3 -- 5$\sigma$ sources which lie within 2$\sigma$ positional uncertainty of a 1.4 GHz source. The positional uncertainty was determined by adding the average 1.4 GHz uncertainty (1.1 arcsec) in quadrature with the positional uncertainty of a 3$\sigma$ source. At 2.5 GHz the allowed positional offset is 3.8 arcsec, and for 5.2 and 8.7 GHz it is 2.8 arcsec. Thus, there are  are 71, 18, and 2 sources at 2.5, 5.2 and 8.7 GHz, respectively, which are low SNR high frequency counterparts to 1.4 GHz sources. We include these sources in supplementary catalogues. To test the reliability of the low SNR supplementary catalogues, we shifted the 1.4 GHz positions by 0.5 arcmin and repeated the matching of 3 -- 5$\sigma$ sources. As a result, we estimate that 4/71 (6\%) of the low SNR 2.5 GHz sources are false, and no false low SNR sources were found at 5.2 and 8.7 GHz. 

\section{Catalogues at 2.5, 5.2 and 8.7 GHz}
\label{catalogues}

The main catalogues for the three frequencies are presented in Tables \ref{catalogue25} to \ref{catalogue87}. These are independent, statistically complete and reliable catalogues for each frequency. We also include supplementary catalogues for each frequency (Tables \ref{catalogue25supp} and \ref{catalogue5287supp}). The supplementary catalogues contain low SNR sources which have a priori 1.4 GHz positions. 

A description of the Tables is as follows. 

{\em Column (1)} --- Source ID. The ID numbering for the supplementary catalogues starts from the end of the main catalogues.

{\em Column (2)} --- Right Ascension in J2000.
 
{\em Column (3)} --- One sigma uncertainty of Right Ascension, in arcsec.

{\em Column (4)} --- Declination in J2000.

{\em Column (5)} --- One sigma uncertainty of Declination, in arcsec.

{\em Column (6)} --- Source peak flux density, in mJy. The values given 
here are not corrected for the systematic effects described in Section \ref{imageanalysis}. ATCA fluxes are generally estimated to be accurate to about 10\%.

{\em Column (7)} --- Source integrated flux density, in mJy.

{\em Column (8) and (9)} --- The {\em deconvolved} major and minor axes (FWHM)
                     of the source, $\Theta$, in arcsec. Only given for successfully deconvolved sources. 

{\em Column (10)} --- The {\em deconvolved} position angle (PA, measured from N through E)
                     of the source, in degrees. Only given for successfully deconvolved sources.

{\em Column (11)} --- The signal-to-noise ratio of the detection,
                     calculated as {\em IMSAD} fitted peak/$\sigma_{\rm
                     local}$.

{\em Column (12)} --- Gaussian fit flags : ``2'' refers to poor integrated flux
density, see Section \ref{srcext} for more details.

\section{Combining the ATHDFS Catalogues}
\label{combcat}

Including the 1.4 GHz catalogue from Paper II, we have catalogues at four radio frequencies. The four images have similar sensitivities, but different area coverages. We now combine these independent catalogues into one consolidated catalogue which lists the flux densities at all frequencies for each individual radio source.

We start by matching the positions of the sources in the 1.4 GHz catalogue
(Paper II) against all source positions in the three higher frequency catalogues. The
allowable offset was set to $3\sigma$, i.e. 3 times the positional 
uncertainties in the catalogues (added in quadrature). This results in a total
of 499 individual sources comprising:
\begin{itemize}
\item the 466 1.4 GHz sources, as presented in Paper II,
\item 11 unmatched 2.5 GHz sources,
\item 5 unmatched 5.2 GHz sources,
\item 4 unmatched 8.7 GHz sources.
\end{itemize}

Each of the 486 sources from this first pass was inspected for missed
matches. The simple $3\sigma$ positional cutoff missed bona fide matches 
due to a variety of reasons, including:
\begin{enumerate}
\item the positional errors being smaller than the image
cellsize, and hence the allowable offset was too small, and 
\item individual sources at 1.4 GHz catalogued as multiple sources at higher 
frequencies and resolution.
\end{enumerate}
After inspection we find that only five 2.5 GHz sources and two 8.7 GHz
sources remain un-matched to sources in the 1.4 GHz catalogue. The final consolidated 
catalogue contains 473 individual sources. 

The consolidated catalogue is presented in Table \ref{combcatalog}. It includes information from the supplementary catalogues. Flux density upper limits (assuming point sources) are listed for 1.4 GHz sources without counterparts at the higher frequencies, if they lie in the catalogued region at that frequency. 

Table \ref{combcatalog} is organised as follows. 

{\em Column (1)} --- Source name. 

{\em Column (2)} --- Right Ascension in J2000. 

{\em Column (3)} --- Declination in J2000.

{\em Column (4)} --- Coordinate flag indicating the frequency from which the source
coordinates are obtained. ``L'' indicates 1.4 GHz, ``S'' indicates 2.5 GHz, 
``C'' indicates 5.2 GHz, and ``X'' indicates 8.7 GHz.

{\em Column (5)} --- Source 1.4 GHz flux density, in mJy.  

{\em Column (6)} --- Source 2.5 GHz flux density, in mJy.  Flux densities from low SNR sources in the supplementary catalogue are listed in brackets. A "-" indicates the source lies outside the catalogued region. 

{\em Column (7)} --- Source 5.2 GHz flux density, in mJy.  Flux densities from low SNR sources in the supplementary catalogue are listed in brackets. A "-" indicates the source lies outside the catalogued region. 

{\em Column (8)} --- Source 8.7 GHz flux density, in mJy.  Flux densities from low SNR sources in the supplementary catalogue are listed in brackets. A "-" indicates the source lies outside the catalogued region. 

{\em Column (9)} --- The spectral index between 1.4 and 2.5 GHz, ($S \propto \nu^\alpha$).

{\em Column (10)} --- The spectral index between 2.5 and 5.2 GHz, ($S \propto \nu^\alpha$).

{\em Column (11)} --- The spectral index between 5.2 and 8.7 GHz, ($S \propto \nu^\alpha$).

{\em Column (12)} --- AGN flag, see Section \ref{alphasect}.

Note that the flux densities in this catalogue are peak flux densities for
sources undeconvolved at that particular frequency, otherwise 
integrated flux densities are presented. The radio spectral indices are discussed in the following section. 

\section{Radio Spectral Indices}
\label{alphasect}

Spectral indices for our sources cannot be obtained from
the original catalogues since the resolution of the images differ between 
the frequencies. In particular, low surface brightness regions that are 
detected in the low resolution 1.4 GHz image may be resolved out at higher 
frequencies. 

We therefore produced low resolution images at 2.5, 5.2 and 8.7 GHz 
by imaging only the {\em uv} data that was equal to, or less than, the 
highest resolution {\em uv} point in the 1.4 GHz image ($6 {\rm km} / 20 {\rm cm} = 30 {\rm
k}\lambda$). The resulting images are of similar resolution 
to the 1.4 GHz image. Flux density measurements from these low resolution images were 
made and the resulting spectral indices for sources detected at more than one frequency are 
given in Table \ref{combcatalog}. The upper limit to $\alpha_{1.4 {\rm GHz}}^{2.5 {\rm GHz}}$ is also given for 1.4 GHz sources that lie within the 2.5 GHz catalogued region. 

\begin{figure}
\centering
\includegraphics[width=10cm]{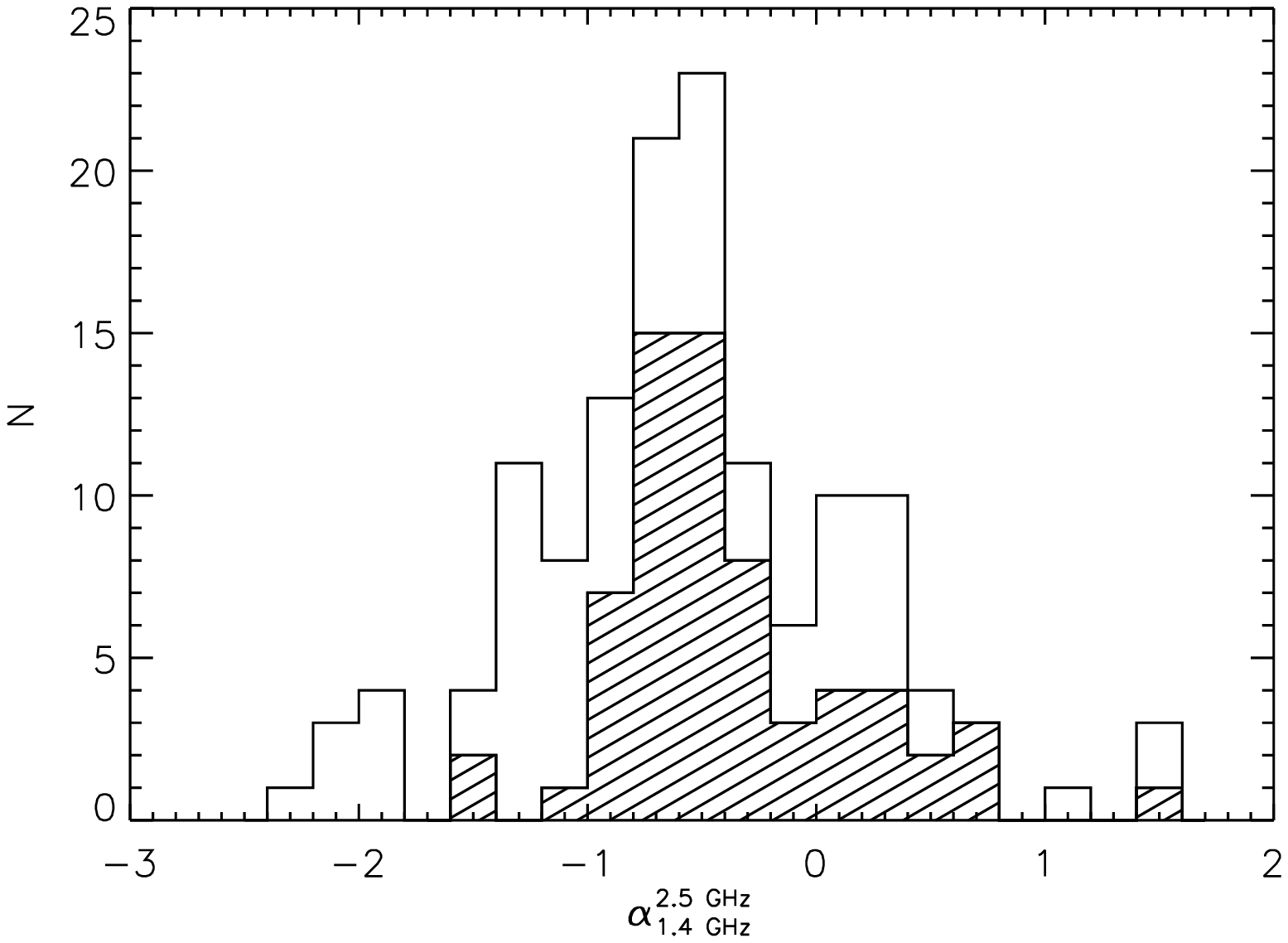}
\caption{Distribution of the spectral index $\alpha_{\rm 1.4 GHz}^{\rm 2.5
GHz}$ of the 136 sources detected at both 1.4 and 2.5~GHz. The hatched histogram shows sources with significant detections in each frequency, while the blank histogram includes the low signal to noise sources from the 2.5 GHz supplementary catalogue.}
\label{alphafig1}
\end{figure}

\begin{figure}
\centering
\includegraphics[width=10cm]{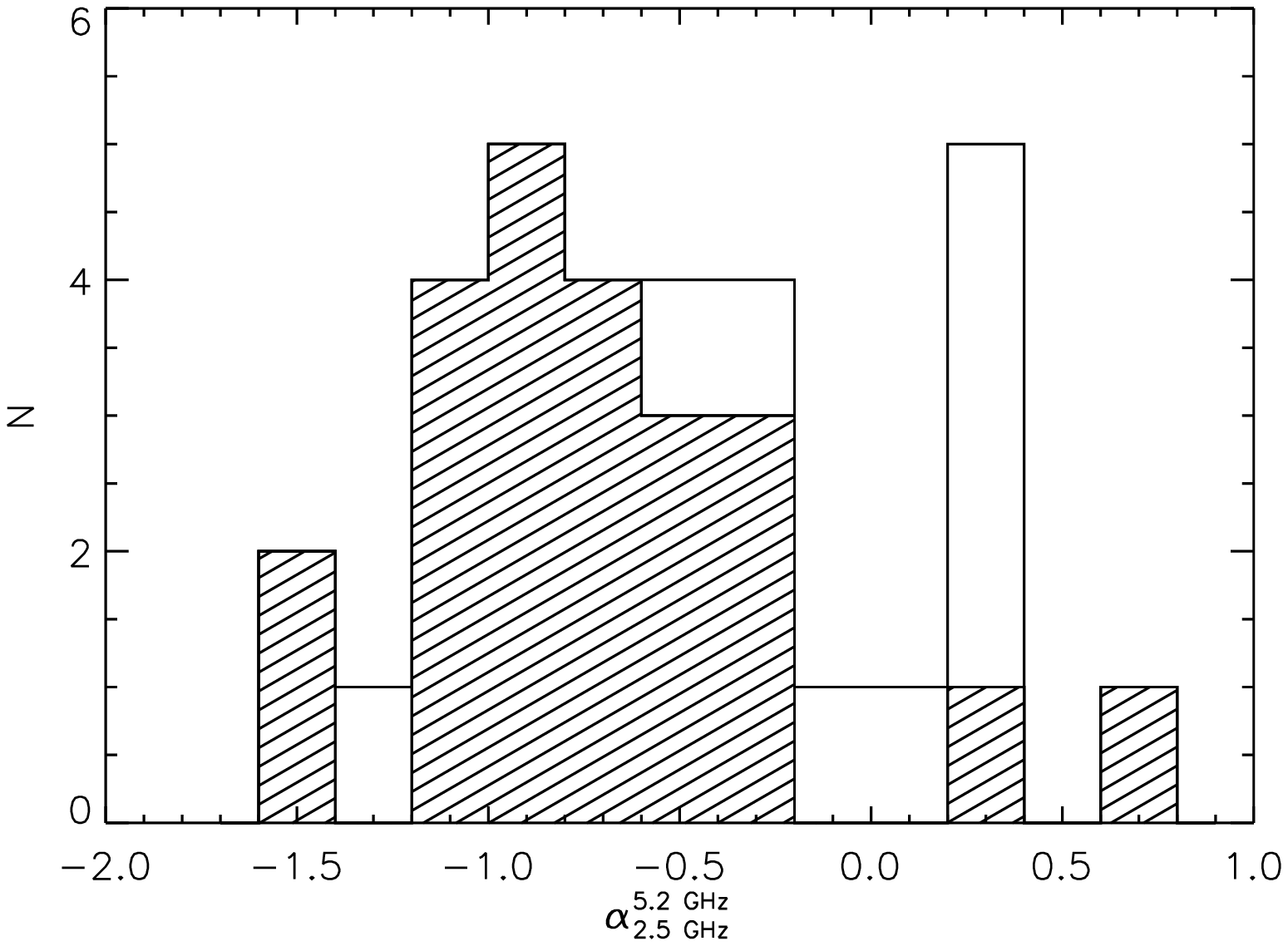}
\caption{Distribution of the spectral index $\alpha_{\rm 2.5 GHz}^{\rm 5.2
GHz}$ of the 32 sources detected at both 2.5 and 5.2~GHz. The hatched histogram shows sources with significant detections in each frequency, while the blank histogram includes the low signal to noise sources from the supplementary catalogues.}
\label{alphafig2}
\end{figure}

\begin{figure}
\centering
\includegraphics[width=10cm]{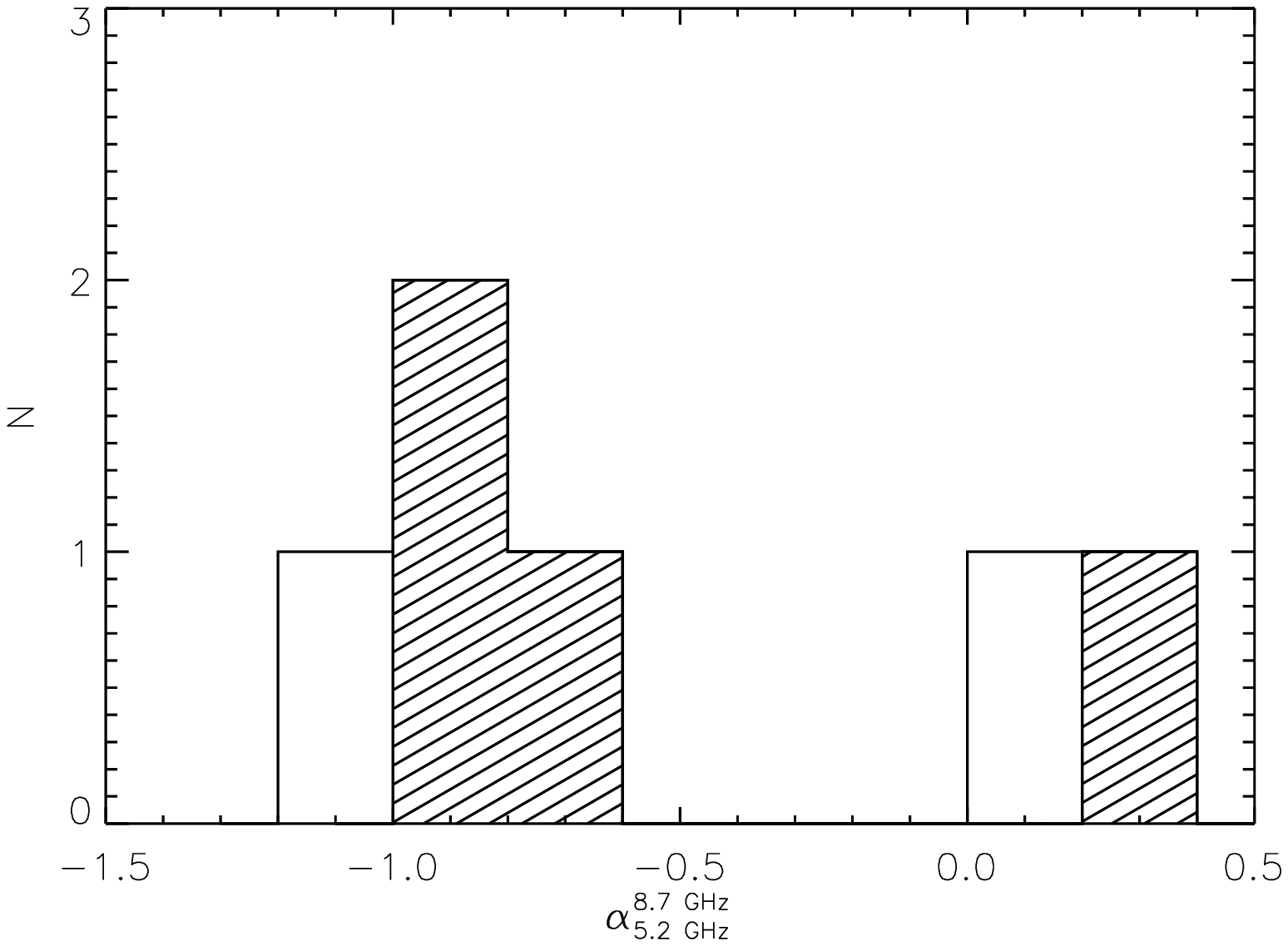}
\caption{Distribution of the spectral index $\alpha_{\rm 5.2 GHz}^{\rm 8.7
GHz}$ of the 6 sources detected at both 5.2 and 8.7~GHz. The hatched histogram shows sources with significant detections in each frequency, while the blank histogram includes the low signal to noise sources from the supplementary catalogues.}
\label{alphafig3}
\end{figure}

\begin{figure}
\centering
\includegraphics[width=12cm]{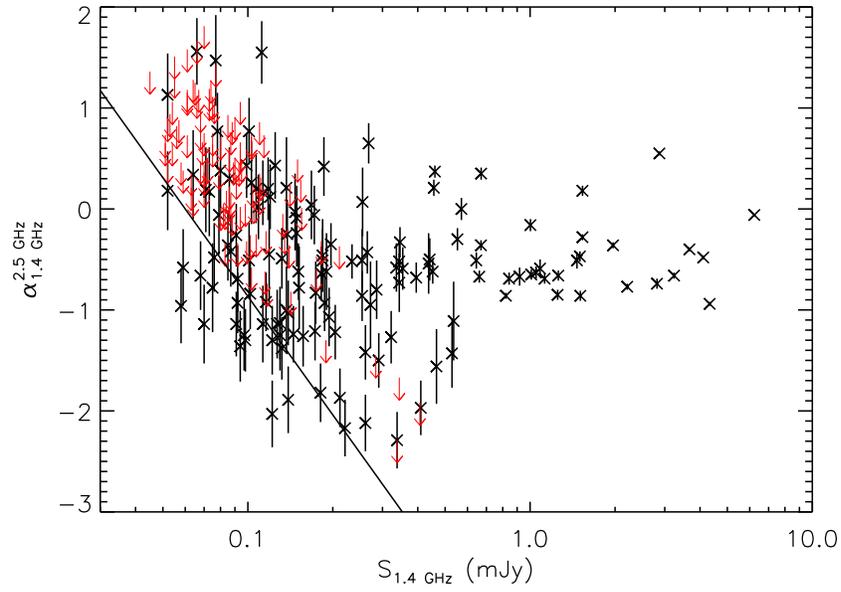}
\caption{The spectral index $\alpha_{\rm 1.4 GHz}^{\rm 2.5
GHz}$ vs 1.4 GHz flux density for the 1.4 GHz sources within the 2.5 GHz catalogued region. The red arrows mark upper limits of sources not detected at 2.5 GHz. The solid line shows the alpha upper limit assuming a detection limit of 0.06 mJy (5.5 $\sigma$) at 2.5 GHz.}
\label{alphafig4}
\end{figure}

The distribution of the spectral indices is plotted in Figures \ref{alphafig1}
to \ref{alphafig3}. The peak in the distribution of the spectral index $\alpha_{1.4 {\rm
GHz}}^{2.5 {\rm GHz}}$  ($S \propto \nu^\alpha$) is $-$0.5. 
This is close to, but flatter than, the canonical value of $-$0.8 expected for radio
synchrotron emission \citep{condon1992}. This is because the spectral indices plotted 
here are only for ATHDFS sources detected at both frequencies, which introduces a bias towards flatter sources. 
Many steep spectrum 1.4 GHz sources are not detected at 2.5 GHz because the images have similar rms 
sensitivities. The alpha upper limit is shown in Figure \ref{alphafig4}, which indicates we can not detect 
faint ($S_{\rm 1.4 GHz} \lesssim 0.06$ mJy) sources with $\alpha_{1.4 {\rm GHz}}^{2.5 {\rm GHz}} \lesssim 0$ at 2.5 GHz.

The mean spectral index $\alpha_{1.4 {\rm GHz}}^{2.5 {\rm GHz}}$ for all sources including the supplementary 2.5 GHz detections is $-$0.54 $\pm$ 0.25. The mean spectral index of the 2.5 GHz selected sample ($S_{\rm 2.5 GHz} > 5.5 \sigma$) is $-$0.39 $\pm$ 0.17. This is consistent with previous work which finds that samples selected at higher frequency have flatter spectral indices (e.g \citealp{windhorst1993, prandoni2006}). 

In Figure \ref{alphafig4} we show the spectral indexes as a function of 1.4 GHz flux density for 136 sources in both the 1.4 GHz and 2.5 GHz catalogues. Our sample is complete to $\sim$0.1 mJy at both 1.4 GHz and 2.5 GHz and the mean spectral index, $\alpha_{1.4 {\rm GHz}}^{2.5 {\rm GHz}}$, for sources with 0.1 $< S_{\rm 1.4 GHz} < 1$ mJy is $-$0.66 $\pm$ 0.28, with a median value of $-$0.62. These values are consistent with the spectral indexes for bright samples at these flux densities
(although between 1.4 and 5 GHz) (Prandoni et al. 2006). Whilst some authors have observed a
flattening of the spectral index at the sub-mJy level \citep{windhorst1993, prandoni2006}, we have limited statistics to explore this regime.

\subsection{Flat and Inverted Spectrum Sources}

The spectral shape of a radio source can give clues to the physical source of
the radio emission. For a homogeneous, optically thin synchrotron radio source with 
constant magnetic field strength $B$, the electron energies have a power law 
distribution of the form:
\[ N(E) \; dE = N_{O} \; E^{s} \; dE \;,\]
and the spectral index $\alpha = (s+1)/2$. The typical observed value of the
spectral index is $<\alpha> = -0.7$, so $<s> = -2.4$, although this only applies at
higher frequencies. At low frequencies the emitting gas is optically thick
and self-absorption becomes important. The radio spectra turns over
to yield $S \propto \nu^{2.5}$, but the exact frequency of the turnover
depends on many parameters, including the magnetic field strength $B$ and
electron density. In AGN, the spectral indices are close to flat but 
progressively steeper ($\alpha < -1$) at higher radio frequencies.
The flatness of the radio spectrum in AGN is attributed to the complex source structure,
where the low frequency turnover is different for various components of the
radio source (e.g. NRAO140; \citealp{marscher88}). When the turnover
frequency is high enough, it is possible to observe an inverted spectrum. 

This flat or inverted spectrum cannot be produced by normal synchrotron emission from star
formation processes. The spectral index of our radio sources therefore
provides us with a diagnostic to identify AGNs in our sample. The uncertainty in the spectral index for faint sources becomes large at about $S_{\rm 1.4 GHz} < 0.5$ mJy (see Figure \ref{alphafig4}). 
We therefore use the following conservative classification scheme: if $S_{\rm 1.4 GHz} < 0.5$ mJy sources must have $\alpha > 0$ to be classed as AGN,  and brighter sources must have $\alpha > -0.3$. The 34 sources which are classed as AGN in this way are flagged as AGN in Table \ref{combcatalog}. 

Three sources have been flagged as a subclass of AGN
called Gigahertz Peaked Spectrum (GPS), because the spectrum is inverted between
1.4 and 2.5 GHz, but steep at higher frequency. This indicates that these
sources may be young, extremely compact AGN (see, for example
\citealp{odea1991, odea1998}). The three GPS sources are discussed in detail in Section \ref{gpssect}.

\section{Discussion of Interesting Sources}
\label{inter_src}

\subsection{Broadline Emitting Radio Galaxy}

Classical radio galaxies have been classed morphologically into two types \citep{fanaroffriley}: 
\begin{enumerate}
\item Fanaroff Riley class 1 (FRI) radio galaxies are AGN with radio jets that
become more diffuse in regions distant from the galactic nucleus, and 
\item Fanaroff Riley class 2 (FRII) radio galaxies are lobe dominated with bright emitting hot-spots in the outer region of their radio lobes. 
\end{enumerate}
There is a strong correlation of morphology with radio luminosity, as noted by
\cite{fanaroffriley}. For low frequencies (178 MHz), sources with radio luminosities
greater than $P=2 \times 10^{25}$ W Hz$^{-1}$ sr$^{-1}$ are almost all FRII type, and lower
luminosity sources are FRI (\citealp{urry95}, and references therein). At GHz frequencies the
luminosity ranges of the two classes overlap by a few decades. 
According to some unification schemes for radio loud AGN, BL Lac objects are 
FRI radio galaxies with jets aligned along the line of sight, and Flat Spectrum Radio Quasars
(FSRQs) are FRII radio galaxies viewed in a similar way \citep{urry95, jackson99, urry02}. 
Spectroscopically, the host of FRIs usually have weak, if any, narrow emission
lines \citep{hine79, laing94}. Most show a quiescent elliptical galaxy spectrum. 
In this sense, FRIs are thought to have no photo-ionizing 
flux, and the difference between FRIs and FRIIs is due to the different 
accretion processes present in these galaxies (e.g. \citealp{baum1995, reynolds1996}, but see \citealp{cao2004}).

The discovery of a FRI radio structure associated with a strong broad line 
emitting quasar E1821+643 is inconsistent with this paradigm.
E1821+643 is an X-ray selected quasar serendipitously found to be associated 
with an FRI \citep{blundell01}. \cite{blundell01} 
suggest that FRI radio galaxies with quasar hosts have not been found 
previously due to the low sensitivity of older radio surveys, such as 3CRR, 
and the low space density of quasars. It was only after very deep radio 
imaging of this quasar was performed with the VLA that the FRI radio 
structure was revealed.

\begin{figure}
\centering
\includegraphics[width = 12cm]{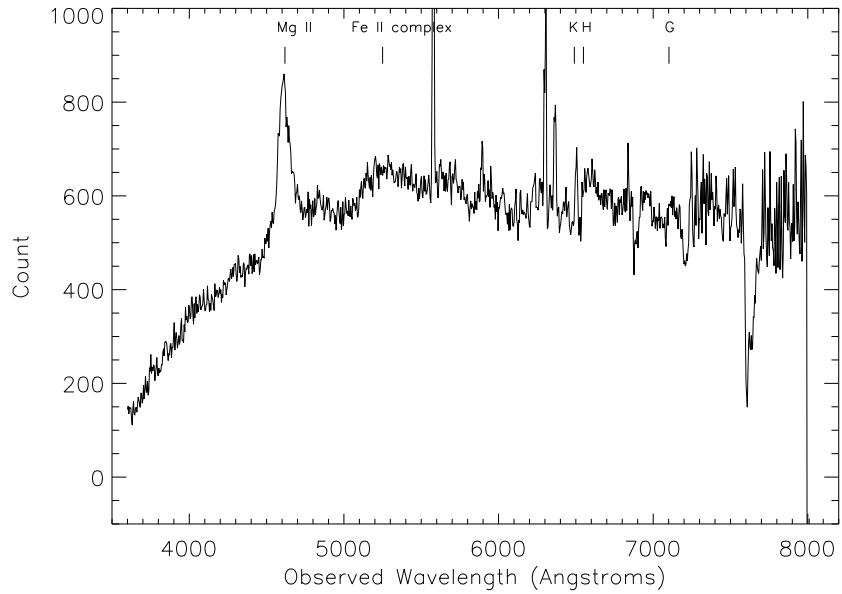}
\caption{Low resolution 2dF optical spectrum of broadline emitting galaxy ATHDFS\_223319.1-604428. The MgII broadline emission and HK absorption features place this galaxy at $z = 0.65$. The strong features in the spectra at about 5600 and 6300 angstroms are artifacts from sky line subtraction.}
\label{fig:fr1spec}
\end{figure}

The deep ATHDFS survey presents a dataset to search for similar objects. 
We have discovered a radio galaxy similar to E1821+643: ATHDFS\_223319.1-604428. 
This radio galaxy is associated with a host galaxy which has optical magnitudes of $V = 19.4$ and $I = 18.1$. 
A low resolution spectrum of this source was taken on the Anglo-Australian
Telescope (AAT) with the 2dF fibre-fed spectrograph, as part of a broader program to obtain spectroscopic redshifts of radio sources
in the ATHDFS. The spectrum obtained of ATHDFS\_223319.1-604428 is shown in Figure \ref{fig:fr1spec}.  We identify the broad feature (5000 km s$^{-1}$) at 4620 $\AA$ as MgII 2799, which is consistent with the H and K Calcium absorption seen in the spectrum, giving a redshift of $z = 0.65$. Table \ref{fr1table} summarises the properties of ATHDFS\_223319.1-604428. The broad MgII line has a restframe measured equivalent width (EW) of 45 $\AA$. This is close to the MgII EW of 50 $\pm$   8 $\AA$ found for  QSOs in the Parkes quarter-Jansky flat-spectrum sample \citep{hook2003}, indicating ATHDFS\_223319.1-604428 has a line emission strength similar to FSRQs. 

\begin{figure}
\centering
\includegraphics[width = 12cm]{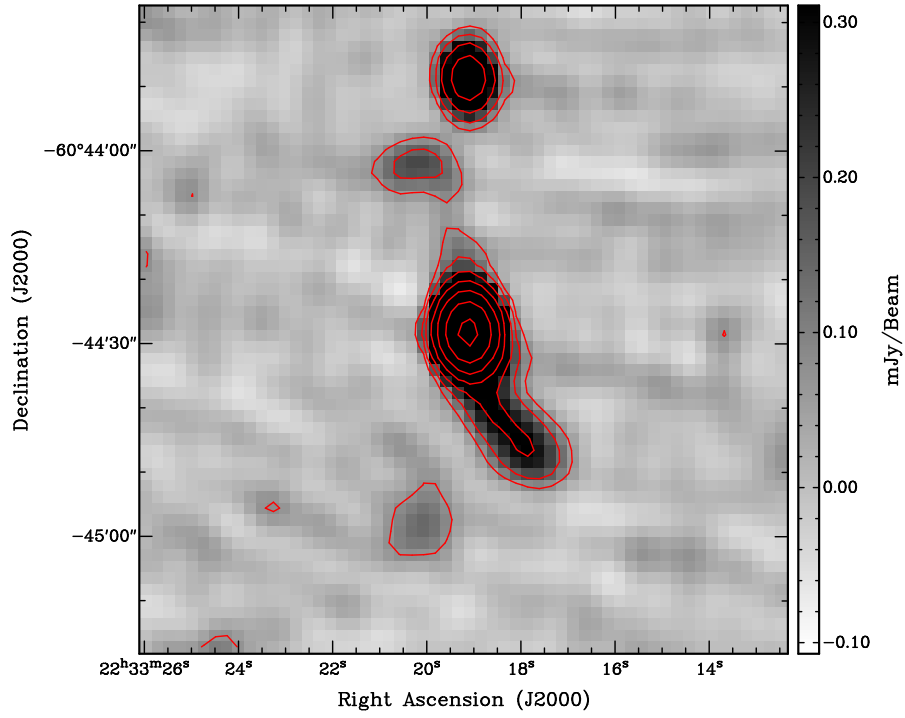}
\caption{Grey scale 1.4 GHz image of ATHDFS\_223319.1-604428, overlaid
with ATHDFS 1.4 GHz contours. The radio contour levels are
set to 3$\sigma$, 6$\sigma$, 12$\sigma$, 24$\sigma$, 48$\sigma$, 96 and 192$\sigma$. }
\label{fig:fr1}
\end{figure}

Figure \ref{fig:fr1} shows the radio morphology of ATHDFS\_223319.1-604428. 
This source is dominated by radio emission from both the core and inner
regions close to the core of the object. On this basis, and having detected weak radio 
lobes (Figure \ref{fig:fr1}), we suggest this is a possible FRI class radio galaxy. 
However, the unresolved source to the north maybe an edge-leading hotspot, so the radio 
morphology is unclear. If the northern source is a hotspot then ATHDFS\_223319.1-604428 
is a moderate redshift FRII with a typical optical host. 

\subsection{Giant Radio Galaxy}

Giant radio galaxies are defined to be radio sources with a projected linear size greater than 1 Mpc. Giant radio galaxies are useful for studying the late stages of radio source evolution because they are thought to be old radio sources. They have been used to test orientation dependent unified schemes and to probe the intergalactic medium at different redshifts (e.g. Subrahmanyan et al. 1996). According to models of radio source evolution (e.g. Kaiser et al. 1997, Blundell et al. 1999), giant radio galaxies must be extremely old (typically older than $10^8$ yr) and usually located in under-dense environments, as compared to smaller radio sources of comparable radio power (e.g. Kaiser \& Alexander 1999).

\begin{figure}
\includegraphics[width = 8cm]{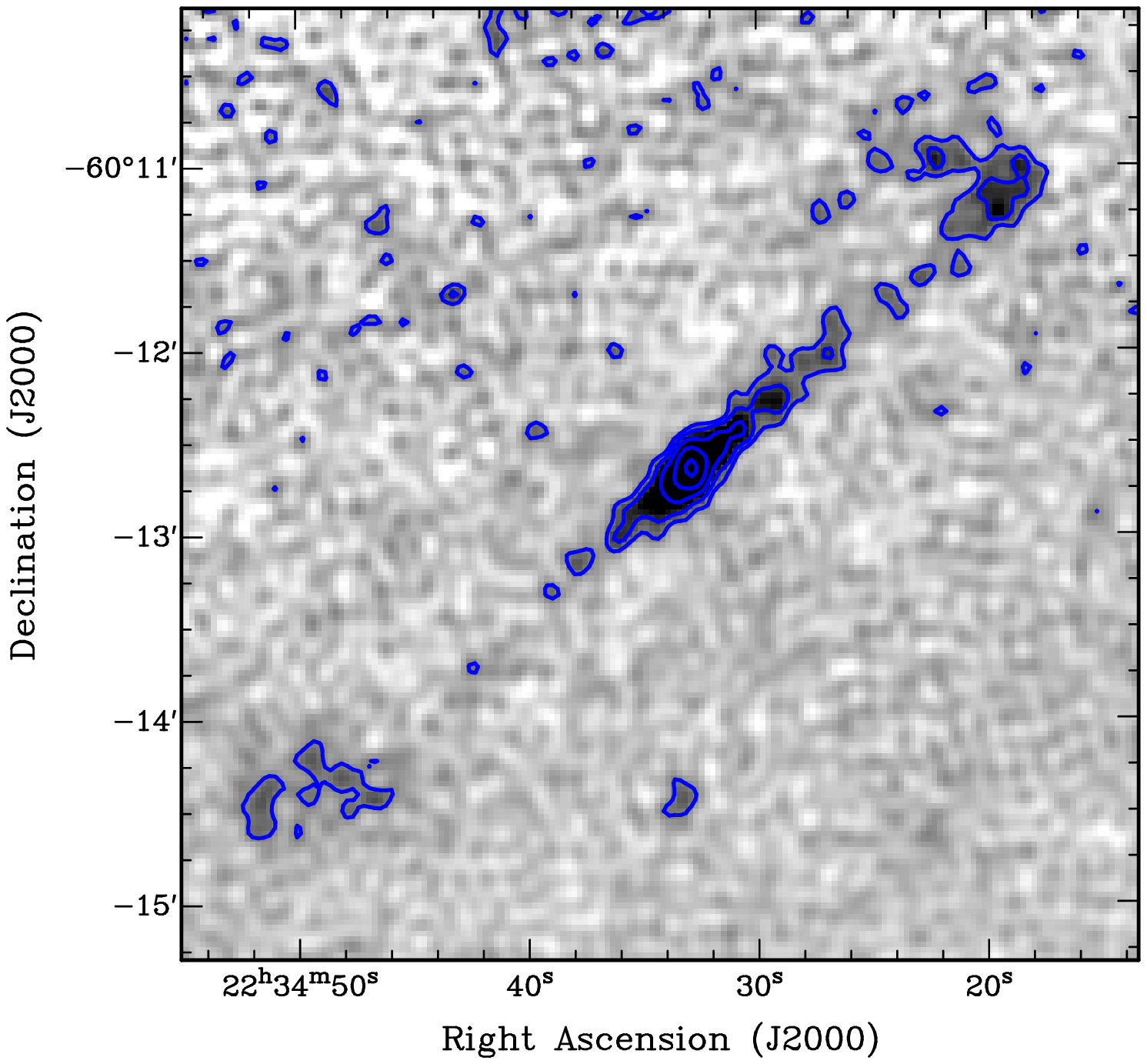}
\includegraphics[width = 9cm]{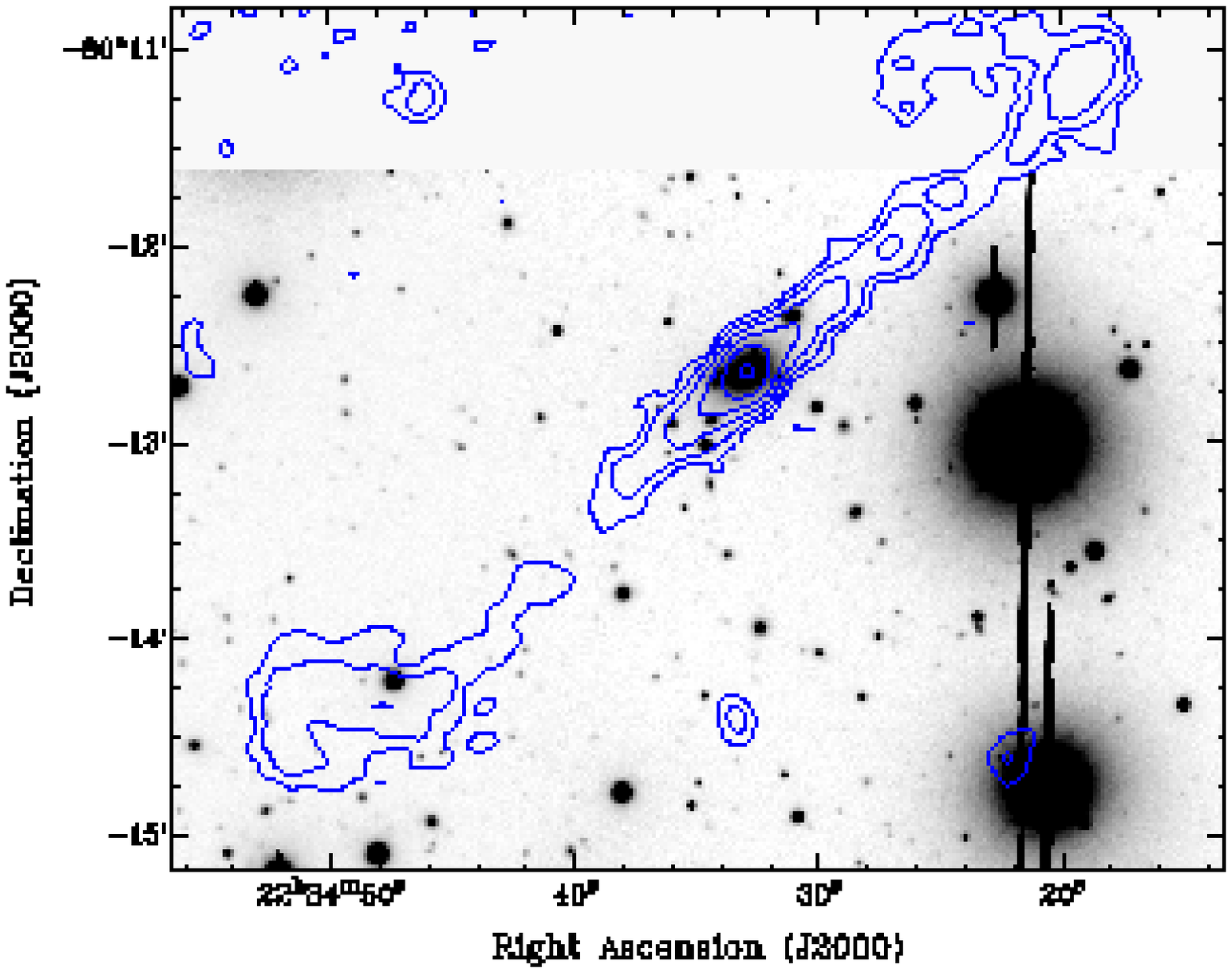}
\caption{{\bf Left:} Grey scale 1.4 GHz image of giant radio galaxy ATHDFS\_223432.9-601239, overlaid with contours from the same image. The radio contour levels are set to 3$\sigma$, 6$\sigma$, 12$\sigma$, 24$\sigma$, and 48$\sigma$. {\bf Right:} Contours from the tapered 1.4 GHz image on CTIO 4m Big Throughput Camera (BTC) I band optical image, showing the bright elliptical host galaxy. Low surface brightness regions of the lobe are more prominent in this tapered radio image. The radio contour levels are set to 3$\sigma$, 6$\sigma$, 12$\sigma$, 24$\sigma$, and 48$\sigma$.}
\label{fig:giantradio}
\end{figure}

Visually inspecting the 1.4 GHz image, we find a giant radio galaxy in the Hubble Deep Field South region which was not included in the original ATHDFS 1.4 GHz catalogue. This is because it lies 27 arcmin from the image centre, and the original catalogue only included sources out to 20 arcmin (see Paper II). This giant radio galaxy, ATHDFS\_223432.9-601239, has a fitted peak of $S_{\rm 1.4 GHz}=$ 2.8 $\pm$ 0.2 mJy and the core has an integrated flux density of 16.4 $\pm$ 1.2 mJy. Some of the extended radio structure maybe resolved out in the full resolution 1.4 GHz image, so we produced a tapered image and find the integrated flux density in the low resolution image is 21 mJy. Contours of this source in both the full and low resolution image are shown in Figure \ref{fig:giantradio}. 

\begin{figure}
\centering
\includegraphics[width = 12cm]{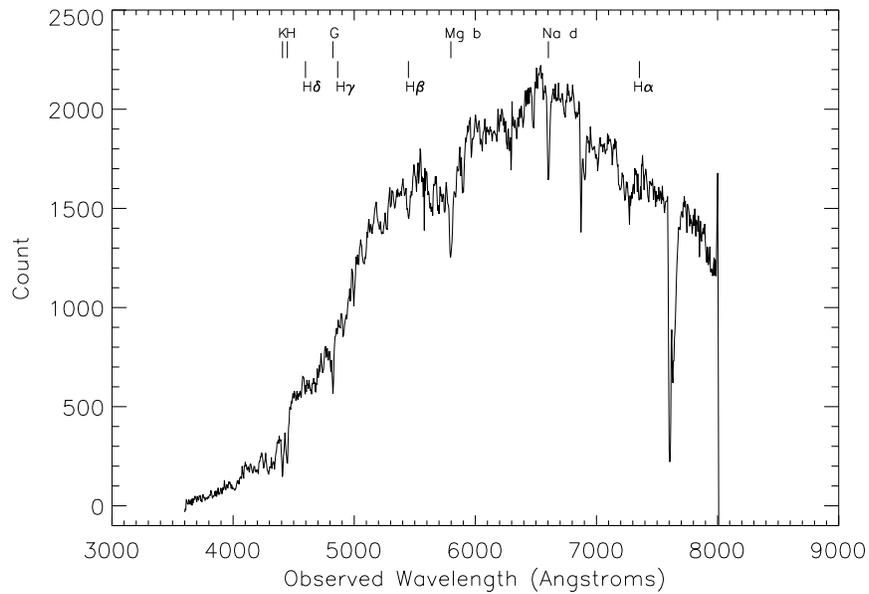}
\caption{Low resolution 2dF optical spectrum of giant radio galaxy ATHDFS\_223432.9-601239. The H and K Ca II, G  band, Mg and Na absorption features place this galaxy at $z = 0.121$. There is no significant H$\alpha$ or nebular line emission, and the higher order Balmer lines are in absorption. This implies that this galaxy has an old stellar population and there is no significant star formation taking place.}
\label{fig:giantradiospec}
\end{figure}

The optical host galaxy is bright, with B mag =  17.68 and I mag = 14.95 \citep{teplitz01}. We summarise the properties of this giant radio galaxy in Table \ref{giantradiotable}. A low resolution spectrum was taken of this object with the 2dF spectrograph in 2001. The spectrum is clearly that of a quiescent elliptical galaxy (Figure \ref{fig:giantradiospec}), as expected of a galaxy with an old stellar population. Fitting the absorption features, we find that this optical host is at a redshift of  0.121.  At this redshift, the linear extent of 390 arcsec translates to a size of 0.83 Mpc and it has a 1.4 GHz radio power of 10$^{23.8}$ W Hz$^{-1}$.

The core and lobes of this giant radio galaxy are individually detected at 853 MHz in the Sydney University Molonglo Sky Survey (SUMSS) survey \citep{bock99}. The core has an integrated flux density of 16.3 mJy at 853 MHz \citep{mauch03}. Hence, the source has a inverted spectral index $\alpha^{\rm 1.4 GHz}_{\rm 853 GHz}$ = 0.51 $\pm$ 0.21, as expected of an AGN.

\subsection{Gigahertz Peaked Spectrum Sources}
\label{gpssect}

Gigahertz peaked spectrum (GPS) sources are compact, powerful radio sources with a well-defined peak in their spectra at around 1 GHz. GPS radio sources make up a significant fraction of the bright radio source population ($\sim$ 10\%), but their contribution at low flux densities is not well known. The turnover in the SED of GPS sources at about a GHz is due to either synchrotron self-absorption \citep{mutel85}, free-free absorption \citep{bicknell97} or absorption from induced Compton scattering \citep{kuncic98}.

GPS sources have linear sizes comparable to or smaller than their optical host galaxies. High resolution radio interferometeric  observations have revealed that GPS sources have double lobed morphology, like classical radio galaxies, but on smaller scales of $\lesssim$ 1 kpc (e.g. Stanghellini et al. 1997). Two explanations have been proposed to explain the compactness of these objects: 1) GPS sources are young unevolved radio sources (e.g. Fanti et al. 1995), and 2) GPS sources are confined by the interstellar medium of the host galaxy (e.g. O'Dea et al. 1991). The current evidence favours the youth scenario: there is no compelling evidence to suggest there is enough dense gas to confine GPS sources (Pihlstrom et al. 2003), the observed expansion speed of GPS hotspots (0.1 -- 0.2$c$) imply ages of $10^3$ years if the lobes expand at constant velocity, and the synchrotron cooling time of the radiating electrons indicate ages of $10^3$ -- $10^5$ years (Murgia et al. 1999). 

In the ATHDFS we have identified three GPS radio sources, ATHDFS\_223327.6-603414, ATHDFS\_223323.2-603249. and ATHDFS\_223259.5-602810. These sources have an inverted SED between 1.4 and 2.5 GHz, and a steep SED at frequencies between 2.5 and 8.7 GHz. We do not have a detailed radio SED for these objects, but our observations constrain the turnover frequency to between 2 -- 4 GHz. These sources are compact, but marginally resolved in the 2.5 GHz image, where they are detected at the highest SNR. They are unresolved in the higher resolution 5.2 and 8.7 GHz images, so they are probably not resolved at 2.5 GHz, but just fit with a slightly larger than point source Gaussian. 

One GPS source, ATHDFS\_223259.5-602810, is not detected in ground-based CTIO imaging to I mag $\sim$ 23 \citep{teplitz01}, but this source lies near a bright star so optical photometry is uncertain. The other two sources lie within the HST flanking field observations, but are not detected. The HST flanking field observations reach a depth of I (F814W) mag = 26.0 \citep{lucas03}. Assuming the Hubble diagram found for GPS sources (\citealp{odea1998}, \citealp{drake04}) holds for these two GPS sources, a radio source with I mag = 26.0 is expected to lie at  $z \gtrsim 2$. Assuming I mag $>$ 26.0 and neglecting K corrections, these sources are fainter than an absolute I (F814W) mag = $-$20.0 at a redshift of $z = 2$. Results from the Sloan Digital Sky Survey (SDSS) have found that $M^*_i$ = $-$21.26 \citep{blanton01}, which corresponds to $M^*_I$ = -21.77 assuming the the conversion from the SDSS team \footnote{http://www.sdss.org/dr4/algorithms/sdssUBVRITransform.html}. GPS sources at redshifts $z < 1$ are usually one to two magnitudes brighter than $M^*$ (\citealp{odea1998}, \citealp{drake04}). This implies these two GPS sources are either unusually faint in the optical, or are at $z \geq  2$. 

\subsection{Ultra-Steep Spectrum Radio Sources}

Tracing radio galaxies out to high redshift provides a way to study the formation and evolution of massive galaxies in the early universe. The well known Hubble $K$--$z$ relation (e.g. \citealp{lilly1984, debreuck2002}) is evidence that high z radio galaxies can be used as a tracer of the most massive star forming galaxies. Radio galaxies have now been found out to redshifts of 5.2 \citep{debreuck2000}. One of the most successful ways to select high redshift radio galaxies (HzRGs) is by using the ultra steep spectrum (USS) selection technique (e.g. \citealp{rottgering1994, blundell1998, debreuck2004}).

We find 29 sources in the ATHDFS with $\alpha^{\rm 2.5 GHz}_{\rm 1.4 GHz}  < -1.1$ (see Table \ref{combcatalog}). Most (21) of the sources have bright (I mag $< 23$) optical counterparts in ground based imaging. Another 4 sources are radio lobes of multiple component radio sources.  The other 4 USS sources could be high redshift AGN. These candidates are listed in Table \ref{uss}. Followup observations of these candidates are required to determine if these sources are HzRGs. 

\section{Summary}

Deep radio observations of the Hubble Deep Field South region with $1\sigma$ rms sensitivities of $\sim$10 $\mu$Jy at 2.5, 5.2 and 8.7 GHz yield catalogues consisting of 71 sources at 2.5 GHz, 24 sources at 5.2 GHz and 6 sources at 8.7 GHz. We have constructed a consolidated radio catalogue from the four frequency observations of the ATHDFS, which has 473 individual radio sources. 

Radio spectral indices were calculated for sources detected at more than one frequency. In deriving the spectral indices, care was taken to correct for spatial resolution effects. For sources detected at both 1.4 GHz and 2.5 GHz, we find the peak in the distribution of the spectral index $\alpha_{\rm 1.4 GHz}^{\rm 2.5 GHz}$ is $-$0.5. We use the spectral index to identify possible AGN. Of the 136 sources with radio spectral information, 34 have flat or inverted spectra and are therefore classified as AGN.

We also identify some interesting radio sources in the ATHDFS survey. One source, ATHDFS\_223319.1-604428, is a radio galaxy at $z = 0.65$ with a broad MgII emission line, and a possible FRI morphology. In the 1.4 GHz image we have also found a giant radio galaxy, ATHDFS\_223432.9-601239,  which has an extent of 0.8 Mpc from lobe to lobe. This giant radio galaxy is at a redshift of $z = 0.121$ and its spectrum is consistent with an old stellar population. Three radio sources have radio SEDs which peak between 2 -- 4 GHz, and hence are classed as Gigahertz Peaked Spectrum sources. Two of these  GPS sources are not detected in HST flanking field observations to I (F814W) mag = 26.0, which implies they are either unusually underluminous  in the optical or lie at high redshifts $z \geq  2$. Using the ultra-steep spectrum selection technique, we find there are 4 possible high redshift radio galaxies in our sample. 

The HDFS has been the target of deep multicolor optical photometry and spectroscopy. The Spitzer Space Telescope has also imaged the HDFS in the infrared. The ATHDFS observations were carried out over several years, and hence it is also a good dataset to investigate time varying radio sources at these frequencies. Future papers in this series will present the optical/infrared identifications of the ATHDFS radio sources and an analysis of time variability. 

\acknowledgements{MTH would like to acknowledge support from an Australian National University PhD Stipend Scholarship. The Australia Telescope Compact Array is part of the Australia Telescope, which is funded by the Commonwealth of Australia for operation as a National Facility managed by the CSIRO.}


\appendix
\section{ATHDFS Combined Catalogue IDs}

Table \ref{athdfs_ids} contains the crossmatch data for the ATHDFS combined catalogue. This table is laid out as follows:

{\em Column (1)} --- Source name. 

{\em Column (2)} --- Source ID in the 1.4 GHz catalogue.

{\em Column (3)} --- Source ID in the 2.5 GHz catalogue. Low SNR sources from the supplementary catalogue are listed in brackets. 

{\em Column (4)} --- Source ID in the 5.2 GHz catalogue.  Low SNR sources from the supplementary catalogue are listed in brackets. 

{\em Column (5)} --- Source ID in the 8.7 GHz catalogue.  Low SNR sources from the supplementary catalogue are listed in brackets.


\clearpage

\clearpage


\begin{table}[hbt]
\small
\centering
\begin{tabular}{cccrcrcr} \hline \hline
$\nu$ & RA & Dec & Hours &  \multicolumn{2}{c}{Synthesised Beam} & 
rms limit & $S_{\min}$ \\
GHz & & & Observed & $b_{\rm maj} \times b_{\rm min}$ & PA & $\mu$Jy & $\mu$Jy \\ \hline
1.4 & 22 33 25.96 & $-$60 38 09.0 & 345 & 7.1'' $\times$ 6.2'' & $-$5.5$^{\circ}$ & 11.0 & $-$168.5\\ 
2.5 & 22 33 25.96 & $-$60 38 09.0 & 94 & 5.1'' $\times$ 4.0'' & 1.7$^{\circ}$ & 10.4 & $-$486.7\\ 
5.2 & 22 32 56.22 & $-$60 33 02.7 & 181 & 3.0'' $\times$ 2.2'' & 10.7$^{\circ}$ & 7.8 & $-$42.3\\ 
8.7 & 22 32 56.22 & $-$60 33 02.7 & 98 & 2.3'' $\times$ 1.7'' & $-$1.6$^{\circ}$ & 11.0 & $-$60.1\\ \hline
\end{tabular}
\caption{Summary of the four frequency ATCA observations centred on the HDFS.}
\label{athdfssummarytab}
\end{table}

\begin{table}
\centering
\begin{tabular}{ccc}\hline\hline
Image   & $a$ & $b$\\ \hline
2.5 GHz & $1.00 \pm 0.01$ & $-0.51 \pm 0.54$ \\
5.2 GHz & $1.00 \pm 0.01$ & $-0.51 \pm 0.47$ \\
8.7 GHz & $1.00 \pm 0.01$ & $-0.53 \pm 0.49$ \\ \hline
\end{tabular}
\caption{Summary of clean bias fit parameters.}
\label{clnbiasfit}
\end{table}

\begin{table}
\centering
\begin{tabular}{cccc}\hline\hline
\multicolumn{4}{c}{2.5 GHz} \\
SNR & sources in  real image & sources in  negative image & reliability (\%)\\ 
$>$5.5 & 71 & 1 & 98.6 \\
5 - 5.5 & 13 & 8 & 38.5 \\
4 - 5 & 171 & 96 & 43.9 \\
3  - 4 & 1300 & 1154 & 11.2 \\ \hline
\multicolumn{4}{c}{5.2 GHz} \\
SNR & sources in real image & sources in negative image & reliability (\%)\\ 
$>5$ & 24 & 1 & 96 \\
4 -  5 & 56 & 35 & 37.5 \\
3 - 4 & 818 & 828 & $\sim$0 \\ \hline
\multicolumn{4}{c}{8.7 GHz} \\
SNR & sources in real image & sources in negative image& reliability (\%)\\ 
$>5$ & 6 & 0 & 100 \\
4 -  5 & 30 & 33 & $\sim$0 \ \\
3 - 4 & 756 & 804 & $\sim$0 \\ \hline
\hline
\end{tabular}
\caption{Summary of reliability estimates.}
\label{reliability}
\end{table}

\begin{deluxetable}{lccccrrrrrrc}
\rotate
\tablewidth{0pt}
\tabletypesize{\small}
\tablecaption{2.5 GHz catalogue.}
\tablehead{\colhead{ID}  & \colhead{RA} & \colhead{$\sigma\alpha$} & \colhead{Dec} & 
\colhead{$\sigma\delta$} & \colhead{$S_{\rm peak}$} &
\colhead{$S_{\rm int}$} & \colhead{$\theta_{maj}$} & \colhead{$\theta_{min}$} & \colhead{PA} & 
\colhead{SN$_{\rm local}$} & \colhead{Flag}}
\startdata
1 & 22:32:26.96 & 0.32 & $-$60:47:46.0 & 0.46 &  0.194 &  0.292 &  3.99 &  2.27 & $-$41.2 &   6.6 & 2 \\
2 & 22:32:54.06 & 0.17 & $-$60:47:37.0 & 0.34 &  0.222 &  0.228 & \nodata & \nodata &  \nodata &   8.2 & 2 \\
3 & 22:33:33.27 & 0.28 & $-$60:46:43.0 & 0.37 &  0.161 &  0.234 &  3.96 &  1.80 & $-$47.4 &   7.8 & 2 \\
4 & 22:33:29.95 & 0.37 & $-$60:45:37.7 & 0.47 &  0.107 &  0.139 &  3.25 &  1.06 &  56.6 &   5.6 & 2 \\
5 & 22:33:42.92 & 0.02 & $-$60:45:24.8 & 0.04 &  1.409 &  1.588 &  2.08 &  1.18 &  11.8 &  65.8 &  \\
6 & 22:34:43.40 & 0.04 & $-$60:44:52.9 & 0.06 &  1.388 &  1.673 &  2.45 &  1.60 & $-$29.3 &  43.3 &  \\
7 & 22:32:10.36 & 0.01 & $-$60:44:33.0 & 0.01 & 26.278 & 43.478 &  4.02 &  3.15 & $-$50.4 & 481.6 &  \\
8 & 22:33:19.13 & 0.01 & $-$60:44:28.3 & 0.01 &  5.222 &  5.969 &  2.23 &  1.24 &   4.0 & 205.5 &  \\
\enddata
\label{catalogue25}
\end{deluxetable}

\begin{deluxetable}{lccccrrrrrrc}
\rotate
\tablewidth{0pt}
\tabletypesize{\small}
\tablecaption{5.2 GHz catalogue.}
\tablehead{\colhead{ID}  & \colhead{RA} & \colhead{$\sigma\alpha$} & \colhead{Dec} & 
\colhead{$\sigma\delta$} & \colhead{$S_{\rm peak}$} &
\colhead{$S_{\rm int}$} & \colhead{$\theta_{maj}$} & \colhead{$\theta_{min}$} & \colhead{PA} & 
\colhead{SN$_{\rm local}$} & \colhead{Flag}}
\startdata
1 & 22:32:32.84 & 0.08 & $-$60:37:37.1 & 0.19 &  0.216 &  0.280 & \nodata & \nodata &  \nodata &  11.2 &  \\
2 & 22:33:16.56 & 0.11 & $-$60:36:27.4 & 0.18 &  0.141 &  0.204 &  2.12 &  1.22 & $-$35.9 &   9.7 &  \\
3 & 22:32:32.70 & 0.19 & $-$60:35:54.9 & 0.32 &  0.089 &  0.141 &  2.76 &  1.24 &  31.1 &   6.2 &  \\
4 & 22:32:32.26 & 0.20 & $-$60:35:41.4 & 0.39 &  0.076 &  0.115 &  2.79 &  1.05 &  24.7 &   5.4 &  \\
5 & 22:32:24.04 & 0.04 & $-$60:35:37.8 & 0.11 &  0.295 &  0.396 &  2.56 &  0.32 &  $-$7.5 &  18.7 &  \\
6 & 22:32:30.21 & 0.13 & $-$60:35:03.5 & 0.29 &  0.076 &  0.083 & \nodata & \nodata &  \nodata &   5.9 & 2 \\
7 & 22:32:43.32 & 0.23 & $-$60:34:42.5 & 0.30 &  0.047 &  0.053 & \nodata & \nodata &  \nodata &   5.1 & 2 \\
8 & 22:32:45.51 & 0.13 & $-$60:34:19.0 & 0.16 &  0.060 &  0.049 & \nodata & \nodata &  \nodata &   6.9 & 2 \\
\enddata
\label{catalogue49}
\end{deluxetable}

\begin{deluxetable}{lccccrrrrrrc}
\rotate
\tablewidth{0pt}
\tabletypesize{\small}
\tablecaption{8.7 GHz catalogue.}
\tablehead{\colhead{ID} & \colhead{RA} & \colhead{$\sigma\alpha$} & \colhead{Dec} & 
\colhead{$\sigma\delta$} & \colhead{$S_{\rm peak}$} &
\colhead{$S_{\rm int}$} & \colhead{$\theta_{maj}$} & \colhead{$\theta_{min}$} & \colhead{PA} & 
\colhead{SN$_{\rm local}$} & \colhead{Flag}}
\startdata
1 & 22:32:43.47 & 0.06 & $-$60:33:51.0 & 0.20 &  0.114 &  0.103 & \nodata & \nodata &  \nodata &   8.0 & 2 \\
2 & 22:32:58.59 & 0.03 & $-$60:33:46.6 & 0.07 &  0.238 &  0.260 & \nodata & \nodata &  \nodata &  19.4 &  \\
3 & 22:33:06.26 & 0.09 & $-$60:33:07.9 & 0.35 &  0.071 &  0.092 & \nodata & \nodata &  \nodata &   5.6 & 2 \\
4 & 22:32:42.66 & 0.06 & $-$60:32:58.1 & 0.20 &  0.077 &  0.044 & \nodata & \nodata &  \nodata &   5.3 & 2 \\
5 & 22:33:08.61 & 0.08 & $-$60:32:52.3 & 0.19 &  0.092 &  0.074 & \nodata & \nodata &  \nodata &   6.2 & 2 \\
6 & 22:33:23.24 & 0.04 & $-$60:32:48.9 & 0.17 &  0.223 &  0.181 & \nodata & \nodata &  \nodata &   7.9 & 2 \\
\enddata
\label{catalogue87}
\end{deluxetable}

\begin{deluxetable}{lccccrrrrrrc}
\rotate
\tablewidth{0pt}
\tabletypesize{\small}
\tablecaption{2.5 GHz supplementary catalogue. These are low SNR (3 - 5.5$\sigma$) sources with a priori 1.4 GHz positions. The ID numbering starts from the end of the main catalogue.}
\tablehead{\colhead{ID}  & \colhead{RA} & \colhead{$\sigma\alpha$} & \colhead{Dec} & 
\colhead{$\sigma\delta$} & \colhead{$S_{\rm peak}$} &
\colhead{$S_{\rm int}$} & \colhead{$\theta_{maj}$} & \colhead{$\theta_{min}$} & \colhead{PA} & 
\colhead{SN$_{\rm local}$} & \colhead{Flag}}
\startdata
72  &  22:34:00.44 & 0.33 & -60:49:03.2 & 0.63 &  0.117 &  0.096 & \nodata & \nodata &  \nodata &   3.8 & 2 \\
73  &  22:34:07.63 & 0.57 & -60:48:47.8 & 0.57 &  0.085 &  0.073 & \nodata & \nodata &  \nodata &   3.0 & 2 \\
74  &  22:33:37.76 & 0.59 & -60:48:33.1 & 0.85 &  0.091 &  0.113 & \nodata & \nodata &  \nodata &   3.6 & 2 \\
75  &  22:32:55.85 & 0.48 & -60:48:22.8 & 0.81 &  0.103 &  0.178 &  4.69 &  3.12 &  15.6 &   4.3 & 2 \\
76  &  22:33:59.72 & 0.68 & -60:48:04.6 & 0.67 &  0.092 &  0.092 & \nodata & \nodata &  \nodata &   3.4 & 2 \\
77  &  22:32:43.13 & 0.43 & -60:47:58.6 & 0.69 &  0.103 &  0.107 & \nodata & \nodata &  \nodata &   4.0 & 2 \\
78  &  22:34:14.73 & 0.24 & -60:47:54.2 & 0.67 &  0.140 &  0.135 & \nodata & \nodata &  \nodata &   4.4 & 2 \\
79  &  22:33:30.51 & 0.35 & -60:46:23.5 & 0.66 &  0.096 &  0.125 &  3.59 &  1.25 & -23.5 &   4.7 & 2 \\
\enddata
\label{catalogue25supp}
\end{deluxetable}

\begin{table}[hbt]
\small
\centering
\begin{tabular}{lccccrrrrrrc} \hline \hline
ID & Ra & $\sigma\alpha$ & Dec & $\sigma\delta$ & $S_{\rm peak}$ & $S_{\rm int}$ & $\theta_{maj}$ & $\theta_{min}$ & PA & SN$_{\rm local}$ & Flag \\ \hline
 \multicolumn{12}{c}{5.2 GHz}  \\
25  &  22:32:53.08 & 0.85 & -60:35:39.1 & 0.85 &  0.036 &  0.035 & \nodata & \nodata &  \nodata &   3.7 & 2 \\
26  &  22:32:29.22 & 0.30 & -60:34:59.9 & 0.43 &  0.054 &  0.080 &  2.49 &  1.02 &  45.0 &   4.2 & 2 \\
27  &  22:32:31.67 & 0.28 & -60:34:24.3 & 0.58 &  0.039 &  0.059 &  2.49 &  1.23 & -15.4 &   3.5 & 2 \\
28  &  22:33:12.04 & 0.35 & -60:34:17.2 & 0.51 &  0.029 &  0.036 & \nodata & \nodata &  \nodata &   3.3 & 2 \\
29  &  22:32:47.65 & 0.22 & -60:33:36.0 & 0.34 &  0.038 &  0.045 & \nodata & \nodata &  \nodata &   4.4 & 2 \\
30  &  22:33:17.73 & 0.29 & -60:32:35.1 & 0.37 &  0.037 &  0.039 & \nodata & \nodata &  \nodata &   3.7 & 2 \\
\hline
\end{tabular}
\caption{5.2 and 8.7 GHz supplementary catalogues. These are low SNR (3 - 5$\sigma$) sources with a priori 1.4 GHz positions. The ID numbering starts from the end of the main catalogue.}
\label{catalogue5287supp}
\end{table}

\begin{deluxetable}{lccccccccccc}
\rotate
\tablewidth{0pt}
\tabletypesize{\tiny}
\tablecaption{The combined ATHDFS catalogue. Flux densities for sources from the supplementary catalogues are listed in brackets. The spectral indices are calculated from flux densities measured on low resolution images, as described in Section \ref{alphasect}.}
\tablehead{\colhead{source name}  & \colhead{RA} & \colhead{Dec} & \colhead{Coord Flag} &
\colhead{$S_{\rm 1.4 GHz}$} & \colhead{$S_{\rm 2.5GHz}$} & \colhead{$S_{\rm 5.2 GHz}$} & \colhead{$S_{\rm 8.7 GHz}$}  & \colhead{$\alpha_{1.4 {\rm GHz}}^{2.5 {\rm GHz}}$} 
& \colhead{$\alpha_{2.5 {\rm GHz}}^{5.2 {\rm GHz}}$} 
& \colhead{$\alpha_{5.2 {\rm GHz}}^{8.7 {\rm GHz}}$} 
& AGN flag }
\startdata
ATHDFS\_J223303.6$-$605751 & 22:33:03.62 & $-$60:57:51.6 & L & 0.508 & - & - & - & - & - & - & \\
ATHDFS\_J223325.8$-$605729 & 22:33:25.88 & $-$60:57:29.3 & L & 0.162 & - & - & - & - & - & - & \\
ATHDFS\_J223255.6$-$605656 & 22:32:55.62 & $-$60:56:56.5 & L & 0.375 & - & - & - & - & - & - & \\
ATHDFS\_J223246.3$-$605654 & 22:32:46.30 & $-$60:56:54.0 & L & 0.254 & - & - & - & - & - & - & \\
ATHDFS\_J223403.4$-$605640 & 22:34:03.43 & $-$60:56:40.3 & L & 0.210 & - & - & - & - & - & - & \\
ATHDFS\_J223243.8$-$605608A & 22:32:43.80 & $-$60:56:08.8 & L & 0.363 & - & - & - & - & - & - & \\
ATHDFS\_J223243.8$-$605608B & 22:32:44.89 & $-$60:56:03.7 & L & 0.176 & - & - & - & - & - & - & \\
ATHDFS\_J223410.5$-$605545A & 22:34:10.58 & $-$60:55:45.4 & L & 2.106 & - & - & - & - & - & - & \\
\enddata
\label{combcatalog}
\end{deluxetable}

\begin{table}
\small
\centering
\begin{tabular}{lc} \hline \hline
flux density (1.4 GHz) &  6.23 mJy \\
redshift $z$  &  0.65 \\
log 1.4 GHz radio power (W Hz$^{-1}$) &  25.0 \\
V app mag  & 19.4 \\
k corrected $M_V$ & -25.6 \\
linear extent & 420 kpc \\ \hline
\end{tabular}
\caption{Summary of broadline emitting radio galaxy ATHDFS\_223319.1-604428. The k-correction to $M_V$ is from \cite{pence1976}.}
\label{fr1table}
\end{table}

\begin{table}
\small
\centering
\begin{tabular}{lc} \hline \hline
1.4 GHz integrated flux density &   16.4 $\pm$ 1.2 mJy \\
1.4 GHz integrated flux density (tapered image) &   21.0 $\pm$ 1.5 mJy \\
853 MHz integrated flux density  &   16.3 $\pm$ 1.4 mJy \\
$\alpha^{\rm 1.4 GHz}_{\rm 853 MHz}$ & 0.51 $\pm$ 0.21 \\
redshift $z$ &  0.121 \\
log 1.4 GHz radio power (W Hz$^{-1}$) &  23.8 \\
V app mag & 16.31\\
k-corrected $M_V$ & -22.6 \\
U $-$ B colour & 0.38 \\
B $-$ V colour & 1.37 \\
V $-$ R colour & 0.68 \\
R $-$ I colour & 0.68 \\
linear extent & 0.83 Mpc \\ \hline
\end{tabular}
\caption{Summary of giant radio galaxy ATHDFS\_223432.9-601239. The k-correction to $M_V$ is from \cite{pence1976}.}
\label{giantradiotable}
\end{table}

\begin{table}
\small
\centering
\begin{tabular}{cc} \hline \hline
name & $\alpha_{1.4 {\rm GHz}}^{2.5 {\rm GHz}}$ \\ \hline  
ATHDFS\_J223408.1$-$604534 & $-$1.82 $\pm$ 0.29  \\
ATHDFS\_J223348.2$-$604026 & $-$1.25 $\pm$ 0.35 \\
ATHDFS\_J223343.7$-$603651 & $-$1.14 $\pm$ 0.39 \\
ATHDFS\_J223329.1$-$602933 & $-$1.42 $\pm$ 0.27  \\ \hline
\end{tabular}
\caption{High redshift AGN candidates selected using the ultra steep spectrum technique.}
\label{uss}
\end{table}

\begin{deluxetable}{lccccccccccc}
\tablewidth{0pt}
\tabletypesize{\small}
\tablecaption{Match data for the combined ATHDFS catalogue. }
\tablehead{\colhead{source name}  & \colhead{1.4 GHz ID} & \colhead{2.5 GHz ID} & \colhead{5.2 GHz ID} & \colhead{8.7 GHz ID}}
\startdata
ATHDFS\_J223303.6$-$605751 & 1 & - & - & - \\ 
ATHDFS\_J223325.8$-$605729 & 2 & - & - & - \\ 
ATHDFS\_J223255.6$-$605656 & 3 & - & - & - \\ 
ATHDFS\_J223246.3$-$605654 & 4 & - & - & - \\ 
ATHDFS\_J223403.4$-$605640 & 5 & - & - & - \\ 
ATHDFS\_J223243.8$-$605608A & 6 & - & - & - \\ 
ATHDFS\_J223243.8$-$605608B & 7 & - & - & - \\ 
ATHDFS\_J223410.5$-$605545A & 8 & - & - & - \\
\enddata
\label{athdfs_ids}
\end{deluxetable}

\end{document}